\documentclass[pdflatex,sn-mathphys-num]{sn-jnl}% Math and Physical Sciences Numbered Reference Style
\usepackage{graphicx}%
\usepackage{multirow}%
\usepackage{amsmath,amssymb,amsfonts}%
\usepackage{amsthm}%
\usepackage{mathrsfs}%
\usepackage[title]{appendix}%
\usepackage[table]{xcolor}% <- keep ONLY this xcolor
\usepackage{textcomp}%
\usepackage[utf8]{inputenc}
\usepackage{textgreek}
\usepackage{manyfoot}%
\usepackage{booktabs}%
\usepackage{algorithm}%
\usepackage{algorithmicx}%
\usepackage{algpseudocode}%
\usepackage{listings}%
\usepackage{float}

\theoremstyle{thmstyleone}%
\theoremstyle{thmstyletwo}%

\theoremstyle{thmstylethree}%

\begin{document}

\title[Article Title]{Modeling Mode and Departure Time Responses to Congestion Pricing: A Spatial and Behavioral Analysis Using Cross-Nested Logit Model}


\author[1]{\fnm{Mohammad Amin} \sur{Ashena}}\email{}%Amin.ashena@ucalgary.ca}

\author[2]{\fnm{Adam} \sur{Weiss}}\email{}%AdamWeiss3@cunet.carleton.ca}

\author*[1]{\fnm{Jason} \sur{Hawkins}}\email{jfhawkin@ucalgary.ca}

\author[1]{\fnm{Lina} \sur{Kattan}}\email{}%Lkattan@ucalgary.ca}

\affil[1]{\orgdiv{Department of Civil Engineering}, \orgname{University of Calgary}, \orgaddress{\street{2500 University Dr, NW}, \city{Calgary}, \postcode{T2N 1N4}, \state{Alberta}, \country{Canada}}}

\affil[2]{\orgdiv{Department of Civil Engineering}, \orgname{Carleton University}, \orgaddress{\street{1125 Colonel By Dr}, \city{Ottawa}, \postcode{K1S 5B6}, \state{Ontario}, \country{Canada}}}


\abstract{Effective congestion management strategies require a detailed understanding of how travellers respond to different pricing interventions. This paper presents an in-depth analysis of traveller behaviour under congestion pricing scenarios, focusing specifically on mode and departure time decisions. Utilizing stated preference survey data from commuters in Calgary, Canada, three discrete choice models including Multinomial Logit, Nested Logit, and Cross-Nested Logit are developed and compared. Results indicate that the Cross-Nested Logit model provides superior behavioural realism and flexibility by capturing simultaneous substitutions across modes and departure times.

Spatial analysis and elasticity assessments reveal substantial geographic variation in traveller sensitivity to pricing, particularly highlighting stronger responses among commuters travelling to high-demand central locations and during peak travel periods. Further elasticity analyses clarify behavioural patterns, identifying traveller groups with varying degrees of flexibility. Policy analyses underscore the effectiveness of targeted, dynamic tolling, particularly cordon-based pricing combined with time-specific toll adjustments, in reducing congestion levels. Additionally, the findings highlight the necessity of complementary measures, including improved transit services and targeted discounts, to ensure equitable outcomes. The findings offer targeted insights into how specific pricing strategies such as cordon, distance, and travel time-based tolls can be used to influence travel behaviour, reduce peak-period congestion, and guide equitable policy design in urban transportation planning.}




\maketitle

\section{Introduction}\label{sec1}

Traffic congestion represents one of the most pressing challenges facing urban areas globally, significantly impacting economic productivity, environmental sustainability, and overall quality of life. As cities expand and urban populations grow, the strain on existing transportation infrastructures intensifies, necessitating innovative and effective solutions to mitigate congestion \citep{lindsney2001}. Among various policy measures, congestion pricing has gained considerable attention as an economic instrument designed to manage travel demand, reduce traffic volumes during peak periods, and encourage shifts towards more sustainable transportation modes \citep{de2011}.

Congestion pricing schemes vary in implementation, often categorized into cordon-based, distance-based, and time-based pricing. Cordon-based pricing typically involves charging vehicles entering a predefined congested urban area, whereas distance-based pricing applies charges according to the length of travel. Time-based pricing schemes levy fees based on the duration of travel within designated zones or specific peak periods. While these mechanisms show promise in influencing travel behaviours, accurately forecasting their effectiveness depends critically on understanding traveller behavioural responses, which involve complex choices related to both travel mode and departure time \citep{gu2018, simoni2019}.

Traditional modelling frameworks, such as Multinomial Logit (MNL) and Nested Logit (NL) models, have historically been employed to predict these behavioural responses. However, these models often fall short in capturing realistic decision-making processes, particularly because travellers simultaneously evaluate both mode and timing of their trips \citep{wen2001generalized}. To address this limitation, the Cross-Nested Logit (CNL) model has emerged as an advanced discrete choice approach capable of capturing the interdependencies and substitution patterns among various travel alternatives more effectively. By allowing alternatives to simultaneously belong to multiple nests such as mode and departure time nests, the CNL model enhances behavioural realism, providing more reliable predictions for policy analysis \citep{papola2004some}.

This paper aims to evaluate commuter behavioural responses to different congestion pricing strategies using an advanced CNL model. By modelling joint mode and departure time choices, the study contributes to a more nuanced understanding of how travellers respond to pricing incentives. The analysis incorporates empirical data, location-based elasticity estimates, and spatial heatmaps to identify nuanced behavioural and geographic patterns. While Calgary, Canada is used as the case study, due to its clear monocentric structure with a downtown core and surrounding suburban and peripheral zones, the findings are intended to inform broader congestion pricing policies in urban areas considering such schemes. Calgary’s absence of an existing CP program allows this research to simulate potential impacts in a similar context. The analysis employs detailed empirical data, spatial location analysis, elasticity calculations, and geographic heatmap visualizations to identify nuanced spatial and behavioural patterns among travellers. Specifically, the study investigates the effectiveness of cordon-based, distance-based, and time-based pricing mechanisms in influencing mode shifts and departure time adjustments.

By comparing the CNL model with traditional MNL and NL models, this study further assesses the predictive performance improvements offered by the more flexible CNL structure. Ultimately, the research seeks to provide policymakers with robust, evidence-based insights and tailored recommendations for designing congestion pricing policies that effectively address traffic congestion in urban settings similar to Calgary, ensuring both equitable outcomes and enhanced transportation efficiency.

Recent work by Heimgartner and Axhausen (2025) highlights the growing need for behavioural models that capture complex substitution patterns and interdependent decision processes. Their multi-stage stated preference study on telework adoption demonstrates that individuals’ responses to policy or contextual changes emerge from layered constraints, preferences, and higher-order behavioural adjustments, rather than isolated choices. Although their focus differs from congestion pricing, their findings reinforce the methodological importance of flexible model structures and carefully designed SP experiments for accurately representing traveller behaviour \citep{heimgartner2025multimodality}. This perspective aligns with the motivation for employing a CNL modelling framework in the present study to better capture the overlapping correlations across mode and departure-time decisions.

This study offers primary contributions. Methodologically, it empirically validates the CNL model as a more behaviourally aligned approach for jointly modelling mode and departure time choices, comparing its performance with standard MNL and NL models. It also integrates stated preference data, designed using a Bayesian D-efficient framework and customized with real-time travel information, into advanced spatial analysis, generating location-specific elasticity estimates and heatmaps that reveal heterogeneous behavioural patterns across urban space. From a policy perspective, the study provides actionable suggestion for urban decision-makers by identifying where and how congestion pricing strategies are most likely to be effective. Although Calgary is used as the case study, the findings and modelling approach are broadly transferable to other metropolitan regions with similar spatial and institutional conditions.

The remainder of this paper is organized as follows: Section 2 provides background on congestion pricing schemes and discrete choice modelling approaches; Section 3 describes the survey design and data collection process. It also outlines the modelling methodology; Section 4 presents the results and comparative analysis of MNL, NL, and CNL models. It discusses elasticity findings and spatial patterns and offers policy insights; and Section 5 concludes with key implications and future research directions.

\section{Background}\label{sec2}

Traffic congestion remains a persistent challenge in urban environments, contributing to substantial economic, environmental, and social costs. These costs include lost productivity, elevated emissions, increased energy consumption, and adverse health outcomes. Exposure to traffic-related air pollution continues to pose significant risks; for example, approximately 10 million Canadians are regularly exposed to harmful levels of transportation-related pollutants \citep{brauer2013ambient}. Motor vehicles are major contributors to this burden: in the United States, road traffic accounts for nearly 60\% of Nitrogen Oxides (NOx) and about 75\% of Carbon Monoxide (CO) emissions \citep{gately2017urban}. These emissions, together with additional congestion externalities such as delays, crash risk, and noise, are typically not internalized by individual travelers, thereby generating substantial societal costs \citep{ecola2009equity}.

Congestion pricing (CP) has been widely recognized as an effective economic instrument for addressing these externalities. First conceptualized by Vickrey \citep{vickrey1969congestion}, CP imposes a monetary charge on drivers using congested road space, thereby encouraging more efficient allocation of limited network capacity \citep{small2007economics}. Recent empirical and simulation-based research further demonstrates the potential of CP to reduce traffic volumes, reshape travel behavior, and improve environmental outcomes. For instance, multi-agent simulations for New York City show that CP can generate differentiated time-of-day and spatial responses, with central areas exhibiting stronger behavioral changes than peripheral districts \citep{he2021nyc}. Such findings underscore the growing recognition that CP impacts are spatially heterogeneous and context-dependent.

Multiple CP schemes are commonly discussed in both research and practice, each influencing traveler behavior through distinct cost mechanisms.

Cordon-based pricing imposes charges on vehicles crossing a designated boundary surrounding a congested urban core. Longstanding examples such as London and Stockholm demonstrate that cordon charges can substantially reduce vehicle entries, improve air quality, and shift travelers toward public transport \citep{eliasson2009cost, leape2006, schuitema2010}.

Distance-based pricing levies charges proportional to the distance traveled within a priced zone, incentivizing reductions in total vehicle-kilometers and discouraging excessive or circuitous travel \citep{daganzo2015}. Because it scales with trip length, distance-based charging may yield distributional advantages in low-density and suburban contexts relative to other CP mechanisms \citep{banister2002transport}.

Travel time–based pricing charges drivers based on time spent in congested conditions, directly linking fees to network performance. Research shows that such pricing encourages travelers to retime trips, shorten routes, or switch modes, particularly during severe peak conditions \citep{aboudina2016, simoni2019}.

Beyond the direct effect on generalized travel costs, recent studies highlight additional behavioral dimensions relevant to CP design. Stated preference evidence from Beijing indicates that congestion charges and reward strategies both influence travelers’ mode-shift potential, with responses mediated by psychological and attitudinal factors \citep{li2019beijing}. Similarly, experimental work applying cumulative prospect theory demonstrates that reference-dependent valuation of schedule delay meaningfully shapes travelers’ departure time adjustments under a congestion charge \citep{geng2023commuter}. Zhang et al. \citep{zhang2023public} emphasized that CP outcomes are not only behavioral but also social: perceived fairness, built environment characteristics, and air-pollution concerns systematically influence public acceptance of pricing schemes. Together, these studies illustrate that modern CP evaluation requires both behavioral modelling and an appreciation of spatial and perceptual heterogeneity.

\subsection{Mode and Departure Time Choices}\label{subsec2.1}

Understanding how travelers adjust their mode of transportation and departure time is central to evaluating the effectiveness of congestion pricing policies. These decisions are shaped by several interrelated factors, including arrival time flexibility, trip chaining, monetary costs, schedule constraints, habitual behavior, and personal or household preferences. In response to pricing, some travelers choose to pay the toll and maintain existing routines, whereas others modify their departure times, switch modes, alter routes, or consolidate activities to avoid peak-period charges. Longer term adjustments such as job relocation or residential moves, may also occur in the presence of sustained pricing interventions \citep{sparrow2020make}.

A substantial body of research examining departure time adjustments follows from the bottleneck formulation introduced by Vickrey \citep{vickrey1969congestion}, in which commuters balance travel time, congestion delay, schedule delay, and monetary charges to minimize generalized cost \citep{yang1997analysis}. Time-of-day pricing builds directly on this framework by imposing higher charges during peak periods, thereby encouraging travelers with sufficient flexibility to shift to earlier or later time windows. Empirical work continues to affirm that commuters typically exhibit stronger aversion to late arrival than early arrival, leading to asymmetric schedule delay preferences that influence departure time shifts under congestion charges \citep{bajwa2008discrete}. Recent behavioral research has further refined this understanding, demonstrating that reference-dependent valuations of schedule delay and tolls can meaningfully influence temporal adjustments; for example, Geng et al.\citep{geng2023commuter} show that loss aversion in the perception of late arrival increases the likelihood of pre-peak departures when a cordon charge is introduced.

Mode and departure time choices are inherently interdependent. A shift from driving to public transit, for instance, often requires alignment with scheduled departure times or increased buffer time for transfers, while telecommuting and ridesourcing alternatives modify the conventional time–cost trade-off. As a result, numerous studies employ joint modelling frameworks to represent the correlation structure linking these decisions. Early work identified significant cross-correlations between mode and departure time utilities \citep{yang2013cross}, setting the stage for more flexible joint-choice representations. More recent research has tested different hierarchical structures for these combined decisions. Ma et al.\citep{ma2020nested} estimated nested logit models for commuting in city of Xi’an, China and found that specifications with departure time at the upper level outperform those with mode at the top of the hierarchy, suggesting that commuters’ temporal preferences may precede their mode selection.

Additional studies have incorporated heterogeneity in substitution patterns. For the Greater Toronto and Hamilton Area, Hossain et al.\citep{hossain2021latent} proposed a latent class structure in which each class embodies a distinct nest configuration for mode and departure time, and class membership varies endogenously with expected maximum utility. Such approaches reveal that travelers differ not only in their time–mode trade-offs but also in the degree to which they substitute across alternatives in response to congestion or pricing changes.

Closest to the present work is the study by Ding et al.\citep{ding2015cross}, who developed a cross-nested logit model to jointly represent mode and departure time choices in the Washington, D.C. region. Their results showed that allowing alternatives to belong simultaneously to mode and time-based nests provides a more behaviorally realistic account of commuters’ substitution patterns than traditional nested logit structures. Their policy simulations highlighted how cross-nesting captures nuanced shifts across both dimensions when tolls or transit improvements are implemented. Building on these insights, the current study adopts a CNL formulation that generalizes substitution across mode and departure time alternatives and applies it to stated preference data explicitly designed around cordon-based, distance-based, and travel time–based congestion pricing scenarios.

\subsection{Elasticities in Mode and Departure Time}\label{subsec2.2}

Elasticities in mode and departure time quantify the degree to which travelers adjust their behavior in response to changes in travel time, monetary cost, or schedule delay. These elasticities vary across individuals and locations, shaped by commute distance, access to alternatives, socio-demographic characteristics, and the spatial organization of activity centers. Evidence from cross-nested applications in Beijing shows that travelers are generally more willing to adjust departure times than to switch modes or relocate, particularly when increases in travel time or cost are modest \citep{yang2013cross}. That study also documented clear distance-based heterogeneity: short-distance commuters exhibited limited responsiveness, whereas commuters traveling 10–20 km showed substantially stronger elastic reactions to time and cost variations.

Recent research has further illustrated that elasticities are sensitive to contextual and environmental factors. For example, multi-agent simulations for New York City indicate that congestion pricing can elicit highly uneven spatial responses, with central districts displaying stronger demand reductions than peripheral areas, reflecting differences in modal availability and baseline congestion \citep{he2021nyc}. Similarly, stated preference work employing cumulative prospect theory demonstrates that reference-dependent valuation of schedule delay can amplify temporal elasticities, especially for travelers who perceive late arrival as a significant loss \citep{geng2023commuter}. These findings suggest that elasticities are shaped not only by cost structures but also by travelers’ underlying risk and schedule-delay attitudes.

Research on temporal flexibility reveals further complexities. While Bajwa et al.\citep{bajwa2008discrete} found that many commuters are willing to depart earlier to avoid late-arrival penalties, Karlstrom \& Franklin\citep{karlstrom2009behavioral} observed substantial inertia in departure-time behavior, indicating that economic incentives alone may not always produce large shifts. These asymmetries help explain why elasticities can differ markedly between early- and late-departure alternatives, even within the same pricing scenario.

Advanced discrete choice models offer the structure needed to capture these heterogeneous responses. Nested and cross-nested logit models, as well as latent class formulations, have shown that elasticities depend on substitution patterns across both mode and departure time dimensions. For instance, Ma et al.\citep{ma2020nested} demonstrated that nesting departure time above mode yields more plausible elastic responses in Xi’an, while Hossain et al.\citep{hossain2021latent} identified latent classes with distinct time–mode trade-offs and systematically different sensitivities to network changes. Ding et al.\citep{ding2015cross} found that allowing alternatives to share membership across mode and time-based nests generates more nuanced and behaviorally reasonable elasticity estimates than traditional nested structures. These contributions collectively highlight the need for flexible model specifications capable of representing correlated decisions and capturing the heterogeneous elastic responses that emerge in the presence of congestion pricing.

\section{Methodology}\label{sec3}

\subsection{Survey Design and Data Collection}\label{subsec3.1}

This study employed both stated preference (SP) and revealed preference (RP) methods to explore traveler responses to three congestion pricing (CP) schemes: cordon-based, distance-based, and travel time-based pricing. The study area encompassed the Calgary Metropolitan Area (CMA), Alberta, Canada, which had a metropolitan population of approximately 1.66 million residents as of 2024 and spans over 825 square kilometers \citep{AlbertaDashboard2025}.

The survey was designed and administered through the Qualtrics platform and targeted respondents who lived in the CMA, were over 18 years old, worked or studied in Calgary, and commuted more than 2 km to their workplace. These screening criteria ensured that the sample captured relevant commuting behaviour, excluding those who might exclusively walk or bike. A professional market research company oversaw survey distribution, ensuring demographic representation consistent with local census data.

The survey consisted of several sections. The initial portion collected detailed demographic and travel behavior information, including age, income, employment status, household composition, vehicle access, parking options, typical commute origin and destination, and work flexibility. Attitudinal questions assessed perceptions of fairness, willingness to pay, and tolerance for congestion under various CP schemes. The final section introduced SP scenarios involving joint mode and departure time choices, presented across three CP schemes with varying toll rates for early, on-time, and late departures.

To construct the SP component, a Bayesian D-efficient experimental design was implemented using Ngene software \citep{ChoiceMetrics2024}. A total of 24 unique SP scenarios were generated and divided into four blocks, with each respondent completing 6 scenarios. JavaScript integration with the Google Directions API enabled the real-time customization of travel time and cost attributes based on respondents’ inputted origin and destination. This ensured that the hypothetical scenarios reflected realistic travel conditions.

Each SP choice task asked respondents to select from four travel modes (driving, transit, riding as a passenger, or ridesharing) and three departure times (on-time, 30 minutes earlier, or 30 minutes later). Each respondent encountered all three pricing schemes throughout their six SP tasks. A pilot test was conducted with 10\% of the final sample to validate and refine the survey design using updated priors. The final survey took approximately 10–15 minutes to complete.

Data collection occurred between September and October 2023. A total of 2,363 individuals accessed the survey, with 705 valid and complete responses retained for analysis. With each respondent answering six SP tasks, the dataset contained 4,230 total choice observations. This comprehensive approach enabled a robust analysis of how CP schemes influence mode and departure time choices, as well as perceptions of fairness across different user profiles and locations.

\begin{table*}[h]
\caption{Travel attributes used in the choice tasks under different weather conditions, pricing schemes, and departure times.}
\label{tbl:attributes}
\footnotesize
\renewcommand{\arraystretch}{1.3}
\begin{tabular*}{\textwidth}{@{\extracolsep\fill}p{0.14\textwidth}p{0.22\textwidth}p{0.18\textwidth}p{0.15\textwidth}p{0.2\textwidth}}
\toprule
\multicolumn{5}{@{}l}{\textbf{Weather:} Sunny, Snowy, Cold without Snow} \\
\multicolumn{5}{@{}l}{\textbf{Pricing Scheme:} Cordon-based [\$6], Distance-based [\$0.6/km], Travel time-based [\$0.24/min]} \\
\multicolumn{5}{@{}l}{\textbf{Departure Time:} Early, On-time, Late} \\
\midrule
\textbf{Attribute} & \textbf{Transit} & \textbf{Driving} & \textbf{Passenger} & \textbf{Ridesharing} \\
\midrule
Travel time (min) &
[0.9$\times$Optimistic, Best Guess, 1.1$\times$Pessimistic] Google Maps (Early: 0.92, Late: 0.85) &
Google Maps &
Google Maps &
Google Maps \\
\midrule
\shortstack{Wait \& walk\\time (min)} &
(5, 7, 10) + Google wait time &
0 &
(5, 8, 11) &
(3, 5, 7) \\
\midrule
Travel cost (\$) &
As claimed\footnotemark, otherwise: 3.5 &
(0.4, 0.5, 0.6)$\times$Distance &
0.5$\times$Driving cost &
Driving cost + 0.25$\times$Driving toll \\
\midrule
Parking cost (\$) &
0 &
As claimed, otherwise: (7.5, 10, 15) &
0.5$\times$Driving parking cost &
0 \\
\midrule
Toll cost (\$) &
0 &
Cordon, Distance, Time based (Early: 0.75, Late: 0.5) &
0.5$\times$Driving toll &
0 \\
\botrule
\end{tabular*}
\footnotetext{“As claimed” refers to cost information directly stated by the respondent; otherwise, the default values are used.}
\end{table*}

\subsection{Model Specification}\label{subsec3.2}

Discrete choice models were used to analyze respondents’ joint mode and departure time choices. The full choice set comprised twelve alternatives, reflecting the combination of four travel modes (driving, passenger, ridesharing, transit) and three possible departure times (early, on-time, late). Let i denote an alternative in this set. The utility $U_i$ for alternative i is specified as a linear function of observed attributes and socio-demographic variables, plus a stochastic error term that is independently and identically distributed (IID) extreme value type 1. Formally, the representative utility for each alternative $i$ is given by Equation 1:
\begin{align}
U_i =\ 
& \text{ASC}_{\text{Mode, Departure}} 
+ \beta_{\text{Time}, M} \cdot \text{time}_{M,D}
+ \beta_{\text{Cost}} \cdot \text{cost}_M  
+ \beta_{\text{Cost}} \cdot \text{Parking}_M \notag \\
& + \beta_{\text{Wait}} \cdot \text{Wait}_{M,D} 
+ \beta_{\text{Toll}} \cdot \text{Toll}_{M,D}  
+ \beta_{\text{Early/Late}} \cdot \text{Early/Late Departure} \notag \\
& + \sum_{k \in K} \beta_k \cdot x_k
\end{align}
\noindent
where: 
\begin{itemize}
    \item \(\beta_{\text{Time}}, \beta_{\text{Cost}}, \beta_{\text{Park}}, \beta_{\text{Wait}}, \beta_{\text{Toll}}\) are parameters capturing the influence of travel time, travel cost, parking cost, waiting time, and toll cost; 
    \item \(x_k\) represents socio-demographic or contextual factors (e.g., age, income, weather conditions); 
    \item \(\beta_k\) is the parameter associated with factor \(x_k\).
\end{itemize}

\footnotetext{Transit fare = $\frac{1}{40} \times$ [2.5 (Low-Income Seniors), 5.6 (Low-Income Class A), 12.5 (Seniors), 39 (Students or Low-Income Class B), 56 (Low-Income Class C), 112 (Regular Monthly Pass), 0 (Provided by Workplace)]}

All attributes such as driving time, transit time, waiting and walking time, toll costs, parking costs, and sociodemographic variables are computed based on a pivot design that uses real values from respondents’ stated addresses and from the literature. Distinct parameter sets are estimated for each of the twelve alternatives (e.g., Driving-Early, Driving-On-Time, Driving-Late, etc.) by interacting generic parameters with departure time or mode-specific characteristics and by including alternative-specific constants.  

\subsubsection{Multinomial Logit (MNL) Model}\label{subsec3.2.1}
The MNL model serves as a baseline for analyzing mode and departure time choices. It is derived from the random utility maximization framework, assuming that individuals choose the alternative that maximizes their utility. The utility \( U_{ni} \) of alternative $i$ for individual $n$ is specified in Equation 2 as:
\begin{align}
U_{ni} = V_{ni} + \varepsilon_{ni}
\end{align}
where \( V_{ni} \) represents the deterministic component of utility, defined as a linear combination of observed attributes (e.g., travel time, cost), and \( \epsilon_{ni} \) is the random error term, assumed to follow a Type I extreme value distribution \citep{train2009discrete}. The choice probability of alternative $i$ is expressed in Equation 3:
\begin{equation}
P_{ni} = \frac{\exp(V_{ni})}{\sum_{j} \exp(V_{nj})}
\end{equation}
While the MNL model is popular for its simplicity and tractability, it assumes the independence of irrelevant alternatives (IIA), meaning the odds of choosing between any two alternatives remain constant regardless of the presence of other alternatives \citep{train2009discrete}.

\subsubsection{Nested Logit (NL) Model}\label{subsec3.2.2}
Building on the recognition that substitution patterns may not only vary across individuals but also structurally between alternatives, an NL model is specified, grouping alternatives into mode-specific nests (Drive, Passenger, Ridesharing, Transit) with departure times nested within each mode. This structure accounts for potential correlation among departure time choices within the same travel mode, partially addressing the limitations of the MNL model through the introduction of nest-specific scale parameters. However, while the NL model allows for correlation within modes, it does not capture possible cross-correlations across departure times or between different modes.

The NL model addresses potential correlations among alternatives by partitioning them into nests. If alternatives within a nest share unobserved factors (e.g., different departure times for the same mode), the NL model captures this correlation \citep{wen2001generalized}. The utility for alternative $i$ within nest $k$ is expressed in Equation 4:

\begin{equation}
U_{nik} = \boldsymbol{V}_{nik} + \varepsilon_{nik}
\end{equation}
where \( \varepsilon_{nik} \) incorporates a correlation structure within the nest. The Nested Logit model relaxes the IIA assumption by allowing alternatives to share unobserved components through nesting structures. In this framework, each alternative belongs to a nest that reflects similarities in unobserved utility, enabling the model to capture substitution patterns among correlated alternatives more effectively \citep{wen2001generalized}. The choice probability is specified in Equation 5:
\begin{equation}
P_{ni} = P(k \mid n) \cdot P(i \mid k, n)
\end{equation}
Here, \(P(k\mid n)\) is the probability of selecting nest k, and \(P(i\mid k,n)\) represents the conditional probability of choosing alternative $i$ within $k$. The flexibility of the NL model in accommodating intra-nest correlations makes it an improvement over the MNL model for many transportation contexts \citep{hess2012joint}.
 
\begin{figure}[h]
    \centering
    \includegraphics[width=\linewidth]{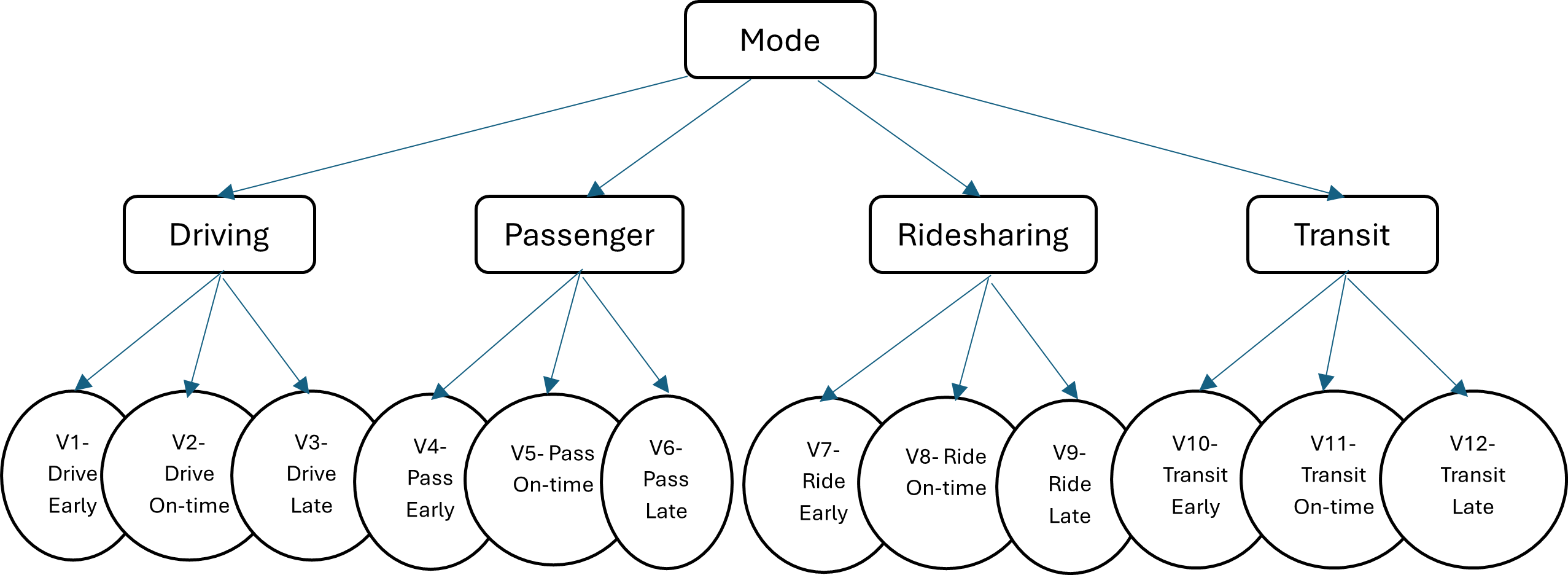}
    \caption{Nested structure of mode and departure time alternatives}
    \label{fig:nested_structure}
\end{figure}
 
The NL model in this case considers travel modes as upper level nests (Drive, Passenger, Ridesharing, Transit) and departure time choices (Early, On-time, Late) as lower level alternatives within each nest as shown in Figure \ref{fig:nested_structure}. This structure captures correlations among departure time options within the same travel mode, reflecting greater similarity among choices associated with the same mode compared to choices across different modes. Nest-specific scale parameters account for the degree of substitution among departure times within each mode.

\subsubsection{Cross-Nested Logit (CNL) Model}\label{subsec3.2.3}

To further enhance the behavioural realism, a CNL model is also estimated, wherein each alternative belongs simultaneously to two nests: a mode nest and a departure time nest. This structure allows for overlapping substitution patterns across both travel modes and departure time options, thereby offering a more flexible representation of choice behaviour under congestion pricing scenarios. The CNL model captures both mode-specific and temporal correlations, improving the model’s ability to represent complex real-world decision-making processes, particularly where travellers may jointly reconsider both when and how they travel in response to tolling interventions. The CNL model extends the NL framework by permitting alternatives to belong to multiple nests with varying degrees of membership. This feature is particularly useful when choices are influenced by overlapping factors such as both mode and departure time \citep{papola2004some}. The utility of an alternative i is represented as it was in Equation 6 with the choice probability defined as:

\begin{equation}
P_{ni} = \frac{\alpha_i \cdot \exp(V_{ni})}{\sum_{j} \alpha_j \cdot \exp(V_{nj})}
\end{equation}

\noindent where $\alpha_i$ denotes the degree of membership of alternative i in its respective nests \citep{bierlaire2006theoretical}. Because it is derived from the Generalized Extreme Value (GEV) framework, the CNL model maintains consistency with random utility theory, offering a tractable closed-form solution while accommodating more complex correlation structures \citep{ding2015cross}.

\begin{table}[h]
\caption{Definition of Alternatives by Mode and Departure Time Nest}
\label{tab:alt_structure}
\small
\renewcommand{\arraystretch}{1.2}
\begin{tabular}{lll}
\toprule
\textbf{Alternative} & \textbf{Mode Nest} & \textbf{Departure Time Nest} \\
\midrule
V1 (Drive, Early)           & Drive        & Early \\
V2 (Drive, On-time)         & Drive        & On-time \\
V3 (Drive, Late)            & Drive        & Late \\
V4 (Passenger, Early)       & Passenger    & Early \\
V5 (Passenger, On-time)     & Passenger    & On-time \\
V6 (Passenger, Late)        & Passenger    & Late \\
V7 (Ridesharing, Early)     & Ridesharing  & Early \\
V8 (Ridesharing, On-time)   & Ridesharing  & On-time \\
V9 (Ridesharing, Late)      & Ridesharing  & Late \\
V10 (Transit, Early)        & Transit      & Early \\
V11 (Transit, On-time)      & Transit      & On-time \\
V12 (Transit, Late)         & Transit      & Late \\
\bottomrule
\end{tabular}
\end{table}

Compared to the simpler MNL and NL approaches, the CNL model offers enhanced flexibility in capturing the multidimensional nature of travel decisions, enabling more accurate policy evaluation. Its effectiveness has been demonstrated in studies assessing congestion pricing impacts, time-varying tolls, and transit improvements \citep{hess2004development, bierlaire2006theoretical}.

\begin{table*}[h]
\caption{Comparison of MNL, NL, and CNL Models}
\label{tab:model_comparison}
\small
\renewcommand{\arraystretch}{1.2}
\begin{tabular*}{\textwidth}{@{\extracolsep\fill}p{0.18\textwidth}p{0.18\textwidth}p{0.27\textwidth}p{0.27\textwidth}}
\toprule
\textbf{Feature} & \textbf{MNL} & \textbf{NL} & \textbf{CNL} \\
\midrule
Correlation handling &
None (IIA holds) &
Mode-based nesting (Drive, Passenger, Ridesharing, Transit) &
Mode and departure time nests simultaneously \\
\midrule
Departure time handling &
Independent &
Nested under mode &
Cross-nested with mode and departure time \\
\midrule
Random taste variation &
No &
No &
No \\
\midrule
Behavioural realism &
Low &
Moderate (structured nesting) &
Very high (flexible cross-nesting) \\
\midrule
Model fit expectation &
Lowest &
Better than MNL &
Best \\
\midrule
Nest scale parameters (\(\mu\)) &
Not applicable &
Significant (shows valid nesting) &
Significant (more complex substitution patterns captured) \\
\botrule
\end{tabular*}
\end{table*}

The CNL model structure, illustrated in Table \ref{tab:alt_structure}, explicitly represents the joint nesting of alternatives across both travel modes and departure-time dimensions. Each alternative simultaneously belongs to two distinct nests: one based on mode choice (Driving, Passenger, Ridesharing, Transit) and another based on departure-time choice (Early, On-time, Late). For instance, the alternative "Drive-Early" (V1) concurrently belongs to the Driving mode nest and the Early departure-time nest, thus reflecting a realistic overlap in how travellers simultaneously evaluate different modes and times when responding to congestion pricing. Table \ref{tab:model_comparison} compares different models that are used in this paper.

 \subsubsection{Elasticity Measures}\label{subsec3.2.4}

An important component of this study involves quantifying how sensitive travelers’ mode and departure time choices are to changes in specific attributes. Elasticities are used to capture these sensitivities in percentage terms. Direct elasticity measures how a change in a particular attribute of an alternative affects the probability of choosing that same alternative, while cross elasticity measures how a change in a particular attribute of one alternative influences the choice probabilities of other alternatives.

For an alternative $i$ and attribute $X$, the direct elasticity of the probability of choosing i with respect to $X_i$ is defined in Equation 7:

\begin{equation}
E_{i,X_i} = \frac{\partial P_i}{\partial X_i} \cdot \frac{X_i}{P_i}
\end{equation}

\noindent where $P_i$ is the probability of choosing alternative $i$. This indicates how much $P_i$ changes, in percentage terms, when $X_i$ changes by 1\%. 

A cross elasticity describes how changing an attribute of one alternative $j$ affects the choice probability of a different alternative $i$ as in Equation 8. Formally,

\begin{equation}
E_{i,X_j} = \frac{\partial P_i}{\partial X_j} \cdot \frac{X_j}{P_i}
\end{equation}

\noindent where $X_j$ is the attribute of alternative $j$. Cross elasticity is crucial for understanding substitution effects, such as how increasing the toll on one mode or route might push travelers toward a different mode or departure time, or how raising transit fares could induce more people to drive. Three key elasticities considered in this study are Travel Time Elasticity, Travel Cost Elasticity and Toll Elasticity.

In the MNL model, elasticities can be derived in closed form due to its mathematical simplicity under the Generalized Extreme Value (GEV) framework \citep{train2009discrete}. While this facilitates efficient computation, the strict IIA assumption may not accurately represent scenarios where alternatives share unobserved similarities. In the CNL model, alternatives can belong to multiple nests with varying degrees of membership \citep{papola2004some, fridstrom2021direct}. Although closed-form expressions remain available under the Generalized Extreme Value (GEV) framework, special attention is needed to account for nest membership parameters ($\alpha_i$) when computing both direct and cross elasticities. This added complexity often yields a more nuanced understanding of how travelers may shift among interrelated alternatives, especially when multiple modes or departure times overlap in attributes.

High direct elasticity for a particular attribute (e.g., toll) suggests that even a modest increase in that attribute could substantially reduce the probability of choosing the associated alternative. Large cross elasticities indicate strong substitution patterns. For example, raising a toll on a congested highway may induce a significant portion of travelers to switch to transit. Also, variations across models (MNL vs. CNL) highlight the importance of correlation structures: if travelers perceive certain modes or departure times as close substitutes, the CNL model will more accurately capture these complexities than the MNL model.

These elasticity measures are instrumental in designing effective congestion pricing strategies. If travelers exhibit high toll elasticity, incremental toll increases in peak periods could meaningfully reduce congestion. To examine how travel time, cost, and toll parameters influence mode and departure time choices, this study estimated both direct and cross elasticities using MNL and CNL models. The CNL model relaxes IIA by allowing alternatives to occupy multiple nests, thus better reflecting how travelers perceive overlaps among options \citep{papola2004some, bierlaire2018calculating}. This enhanced structure provides elasticity estimates that account for the degree of shared attributes between modes or departure times. Consequently, the CNL framework is especially valuable for policy evaluation, as small adjustments to time, cost, or toll attributes in one alternative can lead to notably larger shifts in demand for similar alternatives than suggested by simpler models.

\section{Results and Discussions}\label{sec4}

\subsection{Exploratory Analysis of OD and Routing Patterns}

\begin{figure}[h]
\centering
\includegraphics[width=0.7\textwidth]{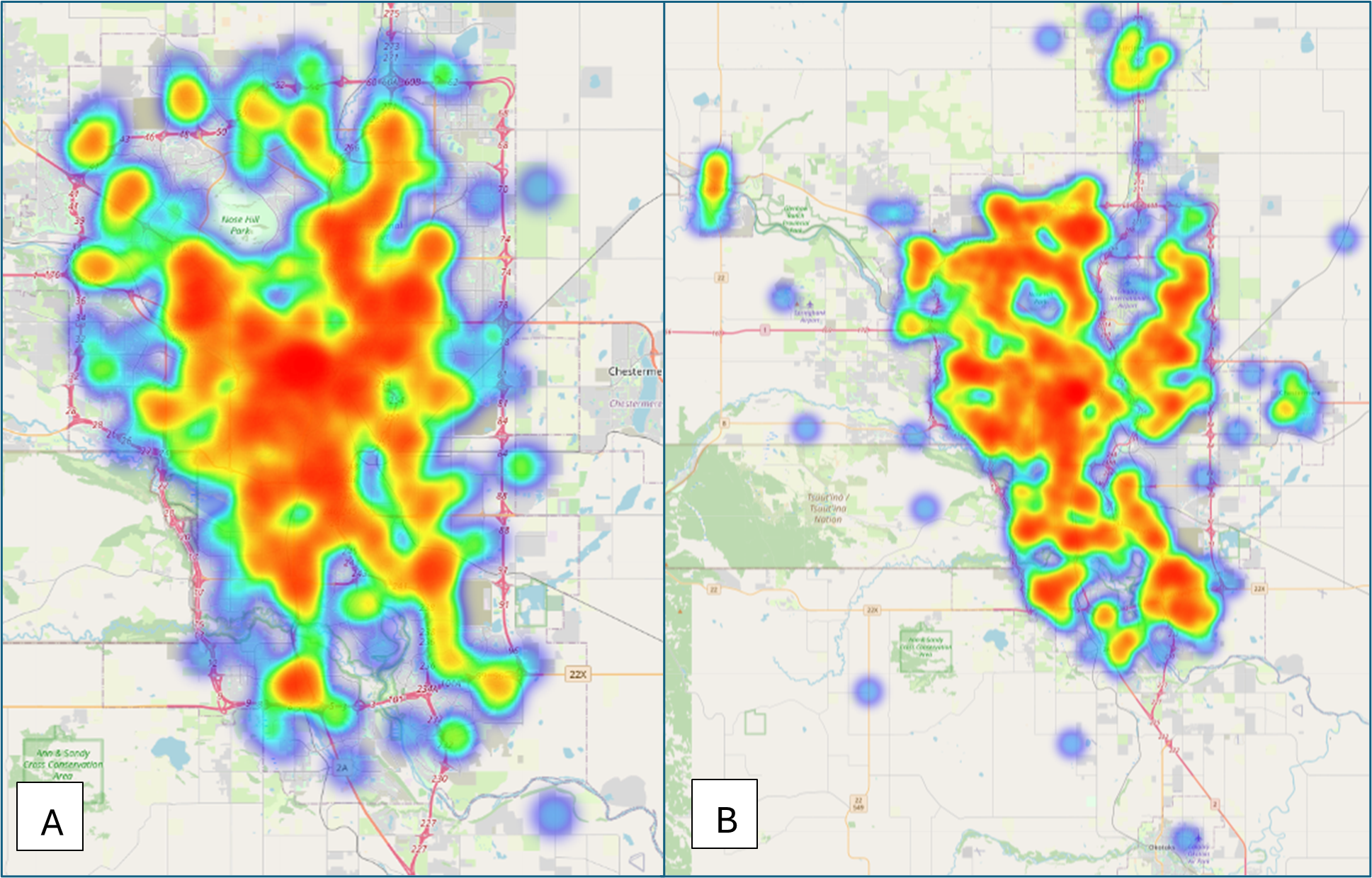}
\caption{Heatmap showing the spatial distribution of respondents’ A) Destination and B) Origin. (Warmer (i.e., red) colors indicate higher concentration)}
\label{fig:O/D_heatmap}
\end{figure}

The spatial analysis of origin and destination points, along with their connecting routes, highlights Calgary's distinctly radial commuting patterns. Trip origins exhibit a broad dispersion across the city, notably concentrated in suburban areas of the northwest (NW), southwest (SW), west (W), northeast (NE), and central residential communities as shown in Figure \ref{fig:O/D_heatmap}. Peripheral origin clusters are evident in adjacent municipalities such as Cochrane (NW), Airdrie (N), Chestermere (E), and Okotoks (S), underscoring significant cross-boundary commuting flows toward Calgary's major employment centers.

In contrast, destinations as shown in Figure \ref{fig:O/D_heatmap} exhibit pronounced centralization, with nearly 20\% of trips concluding within Calgary’s downtown core and the Beltline. This concentration reflects the city's monocentric structure, where employment, commercial, and institutional activities are densely clustered in the core. Beyond this central area, substantial trip destinations align closely with major institutional and employment hubs, including the Foothills Medical Centre and the University of Calgary in the NW, Rockyview General Hospital in the south, the Calgary International Airport in the NE, and industrial employment park in the southeast (SE). While these secondary centers exist, they do not substantially decentralize overall travel demand away from the city center.

Route analysis illustrates pronounced radial travel patterns, reinforcing Calgary’s monocentric urban structure. Major arterials such as Deerfoot Trail, Macleod Trail, and Crowchild Trail accommodate substantial travel volumes from dispersed residential origins towards the densely concentrated central business district. This radial pattern underscores the potential effectiveness of a cordon-based congestion pricing strategy, explicitly targeting central area access points to manage peak-hour congestion efficiently.

The route network clearly demonstrates dense directional flows from peripheral residential areas into central Calgary, particularly from southern, southeastern, and northeastern suburbs. These high-frequency corridors reflect the most congested routes, many of which align closely with existing LRT and BRT transit infrastructure. Notably, the Calgary International Airport in the NE and industrial areas in the SE produce distinct trip patterns contributing significantly to traffic on key arterials outside traditional peak commuting hours.

 \begin{figure}[h]
\centering
\includegraphics[width=0.75\textwidth]{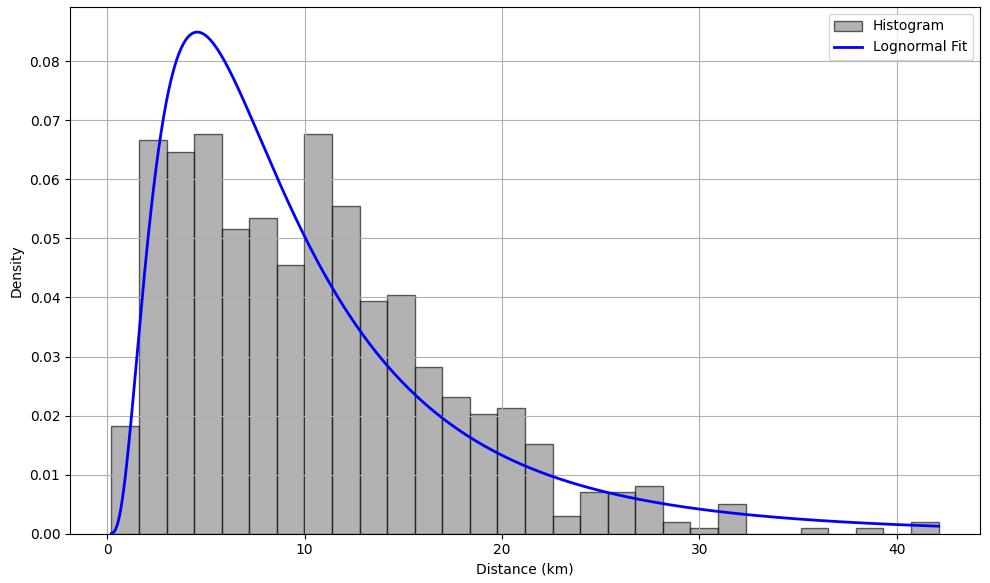}
\caption{Distribution of distances between origin and destination coordinates}
\label{fig:distribution-of-distance}
\end{figure}

Given an average commute distance of around 10 km shown in Figure \ref{fig:distribution-of-distance} which can be fitted to a lognormal distribution $(\mu = 2.11,\ \sigma = 0.77)$, cordon-based congestion pricing emerges as particularly suitable for Calgary’s spatial travel structure. Such a targeted approach would impose costs predominantly on trips that contribute most significantly to central area congestion, rather than dispersing costs across less impactful peripheral trips.

Policy implementation should consider the varying effectiveness of pricing strategies across different regions of the metropolitan area. While cordon-based pricing around the downtown core may be supported by the existing high-frequency transit network, distance-based and time-based schemes affecting peripheral zones require careful planning due to limited access to alternative modes. Effective deployment of these pricing mechanisms depends on concurrent improvements in transit infrastructure and service coverage, particularly in suburban and intermunicipal areas. Coordination with municipalities such as Cochrane, Airdrie, Chestermere, and Okotoks is essential to ensure consistent application and mitigate unintended consequences for commuters reliant on regional road networks. implementation must address equity concerns associated with key destinations such as hospitals, educational institutions, and the airport, where trips involve essential and potentially vulnerable populations. 

While Calgary's monocentric urban form, with dispersed residential origins converging into a dense downtown core, may be structurally compatible with cordon-based congestion pricing, the value of this study lies in its behavioral insights. By modelling joint mode and departure time choices, the analysis reveals nuanced substitution patterns and differential sensitivities among commuter groups. These insights are further explored in the subsequent sections through discrete choice modelling comparisons and elasticity analyses. 

\begin{figure}[h]
    \centering
    \includegraphics[width=0.45\textwidth]{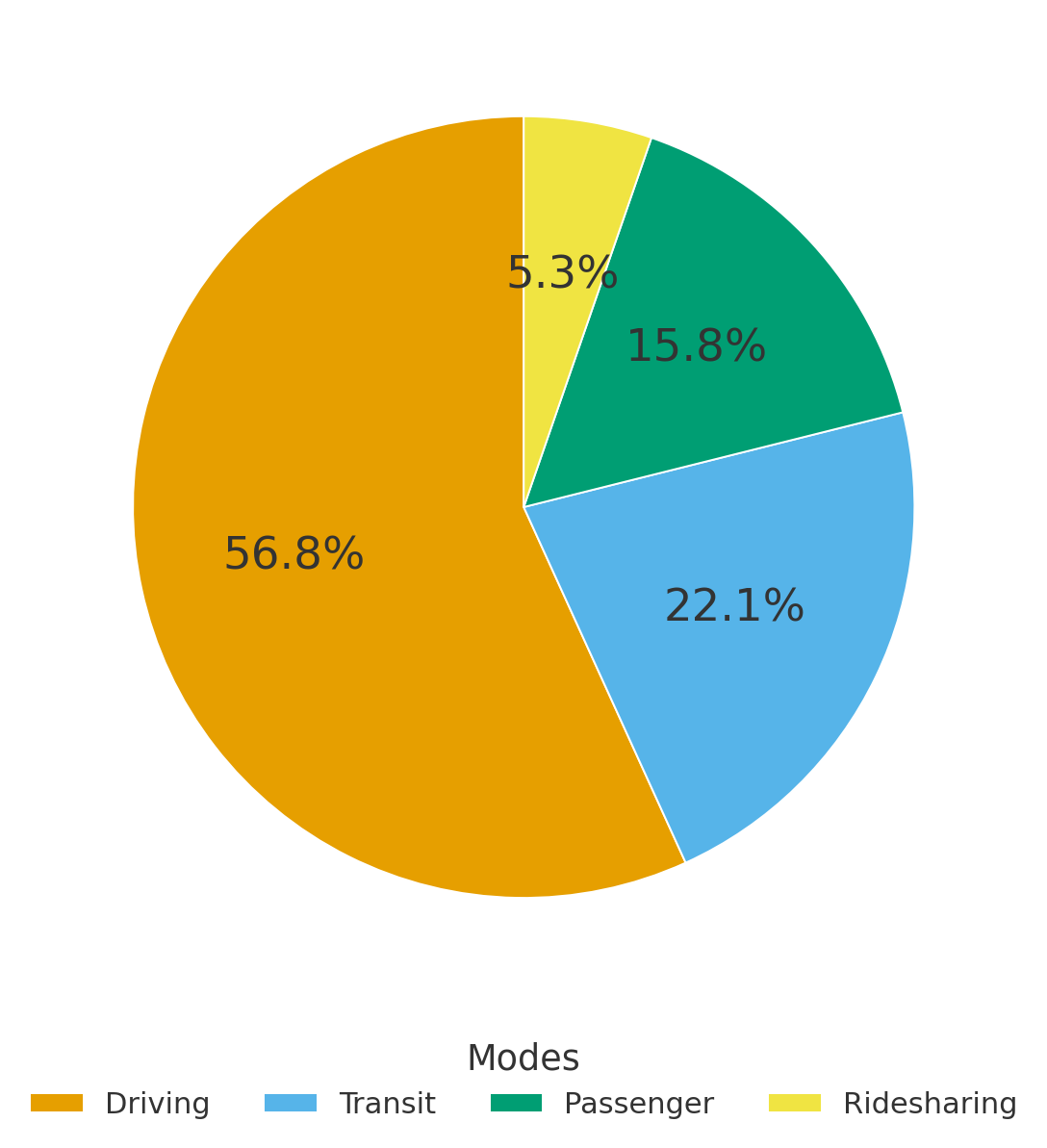}
    \caption{Mode share distribution}
    \label{fig:mode_share}
\end{figure}

The descriptive analysis of stated preference (SP) survey data provides valuable insight into travellers' underlying mode choices and scheduling patterns before any model calibration. Figure \ref{fig:mode_share} illustrates mode shares directly derived from respondents' selections, highlighting a strong preference for driving (56\%), followed by transit, passenger, and ridesharing. This modal split aligns closely with typical commuting patterns in auto-dependent urban areas like Calgary \citep{MacrotrendsCalgary2025}.
 
\begin{figure}[h!]
    \centering
    \includegraphics[width=0.85\textwidth]{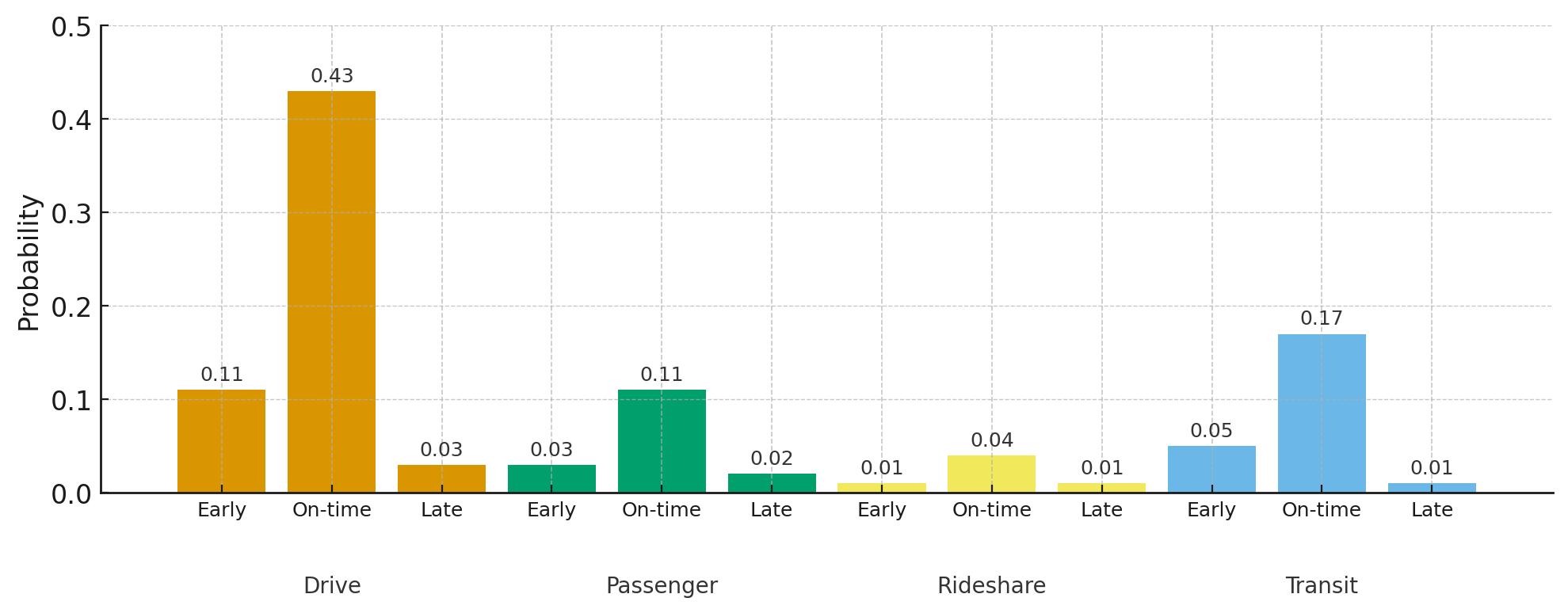}
    \caption{Choice probabilities across all considered alternatives}
    \label{fig:choice_probabilities}
\end{figure}

Figure \ref{fig:choice_probabilities} illustrates choice probabilities across all 12 mode-departure time combinations, revealing more detailed behavioral preferences. Notably, Drive On-time dominates with a probability of 0.43, far exceeding all other options. This suggests a prevailing inertia among commuters to drive and depart during peak periods, despite potential congestion or tolling. Drive Early and Passenger On-time both follow at 0.11, indicating that some travelers either adjust their departure times or opt for shared rides but still align with peak-period commuting.

Transit usage is primarily concentrated around the On-time departure (probability of 0.17), significantly surpassing Transit Early (0.05) and Transit Late (0.01). This distribution suggests transit users typically adhere closely to scheduled services, reflecting limited flexibility in transit schedules. This pattern also aligns with expectations, as transit users are typically not directly affected by congestion pricing scenarios, except when influenced indirectly through substitutions away from driving modes due to pricing schemes.

Ridesharing alternatives exhibit notably low choice probabilities across all departure times, particularly for early and late periods, each below 0.02. These low probabilities could result from limited familiarity, reduced availability, or commuters’ discomfort with sharing rides with unfamiliar individuals. Understanding this baseline traveller behaviour through descriptive statistics aids in interpreting subsequent modelling results, informing how effectively different congestion pricing mechanisms could incentivize mode or departure-time shifts.

\subsection{Modeling Traveler Behavior under Congestion Pricing}

\subsubsection{Multinomial Logit model as the base model}

The MNL model provides initial insights into how travel mode and departure time choices are influenced by individual sociodemographic characteristics, travel attributes, and contextual factors under congestion pricing scenarios. Estimated coefficients as shown in Table \ref{tab:MNL_results} are statistically significant at the 95\% confidence level, and the signs of the parameters are behaviorally plausible.

Sociodemographic factors significantly influence choice behavior: for example, older respondents are more likely to prefer driving and transit, while individuals with university education are less likely to drive. Household size, income level, and vehicle ownership also exhibit expected effects, with high-income individuals showing a preference for late departures and low-income individuals less likely to drive. Travel attributes such as cost, travel time, and waiting time (i.e. for respondents who choose transit and shared rides) are negatively associated with utility, as expected. The toll coefficient is negative and statistically significant, confirming that higher toll charges deter the choice of toll-affected modes. Notably, weather conditions also play a significant role, with snowy conditions increasing the likelihood of early departure and sunny weather increasing the attractiveness of driving. In terms of departure time preferences, the coefficients for scheduled delay costs highlight a strong disutility for early and late departures compared to on-time travel emphasize travelers’ sensitivity to schedule deviations. Finally, policy-related variables such as cordon-based and time-based pricing demonstrate significant effects, suggesting that pricing strategies can effectively influence departure time and mode choice decisions.

Overall, the MNL model establishes a strong foundation for understanding the determinants of travel behaviour under congestion pricing; however, its limitations related to the IIA assumption necessitate the use of more flexible models, such as NL and CNL models to better capture heterogeneity and substitution patterns.

\begin{sidewaystable} 
\caption{Multinomial Logit Model Estimation Results}
\label{tab:MNL_results} 
\small 
\renewcommand{\arraystretch}{1} 
\begin{tabular*}{\textheight}{@{\extracolsep\fill}lllccc} 
\toprule 
\textbf{Parameter Name} & \textbf{Mode} & \textbf{Dep. Time} & \textbf{Value} & \textbf{Std. Err.} & \textbf{t-test} \\ 
\midrule 
Drive ASC & Driving & All & 0.962 & 0.168 & 5.71 \\ 
Ridesharing ASC & Ridesharing & All & -0.762 & 0.222 & -3.44 \\ 
Transit ASC & Transit & All & -0.091 & 0.247 & -0.366 \\ 
Passenger ASC (fixed to zero) & Passenger & All & 0 & -- & -- \\ 
Scheduled Delay Cost for Early Departure ($\beta_{\text{Early}}$) & All modes & Early & -1.51 & 0.064 & -23.4 \\ 
Scheduled Delay Cost for Late Departure ($\beta_{\text{Late}}$) & All modes & Late & -2.64 & 0.09 & -29.3 \\ 
In Vehicle Time for Driving ($\beta_{\text{Time,Car}}$) & Driving & All & -0.026 & 0.011 & -2.28 \\ 
In Vehicle Time for Passenger ($\beta_{\text{Time,Passenger}}$) & Passenger & All & -0.035 & 0.011 & -3.25 \\ 
In Vehicle Time for Ridesharing ($\beta_{\text{Time,Ridesharing}}$) & Ridesharing & All & -0.04 & 0.012 & -3.5 \\ 
In Vehicle Time for Transit ($\beta_{\text{Time,Transit}}$) & Transit & All & -0.021 & 0.004 & -5.63 \\ 
Cost ($\beta_{\text{Cost}}$) & All modes & All & -0.03 & 0.007 & -4.53 \\ 
Waiting and Walking Time ($\beta_{\text{Wait}}$) & \shortstack{Ride, Pass.,\\ Transit} & All & -0.005 & 0.002 & -2.4 \\ 
Toll (Congestion Pricing) ($\beta_{\text{Toll}}$) & Drive, Pass. & All & -0.039 & 0.012 & -3.35 \\ 
Drive for Old Ages ($\beta_{\text{AgeOld,Drive}}$) & Driving & All & 0.88 & 0.265 & 3.32 \\ 
Transit for Old Ages ($\beta_{\text{AgeOld,Transit}}$) & Transit & All & 1.80 & 0.295 & 6.09 \\ 
Early Departure for Young Ages ($\beta_{\text{AgeYoung,Early}}$) & All modes & Early & -0.338 & 0.086 & -3.91 \\ 
Male Passengers ($\beta_{\text{Gender,Passenger}}$) & Passenger & All & -0.361 & 0.088 & -4.1 \\ 
White Race for Driving ($\beta_{\text{RaceWhite,Drive}}$) & Driving & All & 0.253 & 0.076 & 3.36 \\ 
Snowy Weather for Early Departure ($\beta_{\text{Snowy,Early}}$) & All modes & Early & 0.77 & 0.081 & 9.57 \\ 
Snowy Weather for Transit ($\beta_{\text{Snowy,Transit}}$) & Transit & All & -0.881 & 0.19 & -4.64 \\ 
Sunny Weather for Driving ($\beta_{\text{Sunny,Drive}}$) & Driving & All & 0.531 & 0.169 & 3.15 \\ 
Cold Weather for Transit ($\beta_{\text{Cold,Transit}}$) & Transit & All & -0.895 & 0.193 & -4.64 \\ 
University Education for Driving ($\beta_{\text{EducationUniversity,Drive}}$) & Driving & All & -0.406 & 0.076 & -5.37 \\ 
Free Parking for Driving ($\beta_{\text{FreeParking}}$) & Driving & All & 0.378 & 0.1 & 3.78 \\ 
Household size for Driving ($\beta_{\text{Householdsize,Drive}}$) & Driving & All & -0.26 & 0.097 & -2.68 \\ 
Household size for Passenger ($\beta_{\text{Householdsize,Passenger}}$) & Passenger & All & 0.292 & 0.108 & 2.7 \\ 
High Income for Late Departure ($\beta_{\text{IncomeHigh,Late}}$) & All modes & Late & 0.388 & 0.128 & 3.04 \\ 
High Income for Ridesharing ($\beta_{\text{IncomeHigh,Ridesharing}}$) & Ridesharing & All & -0.557 & 0.167 & -3.34 \\ 
Low Income for Driving ($\beta_{\text{IncomeLow,Drive}}$) & Driving & All & -0.447 & 0.158 & -2.83 \\ 
Vehicle Ownership for Driving ($\beta_{\text{VehicleOwnership,Drive}}$) & Driving & All & 0.596 & 0.079 & 7.58 \\ 
No Vehicle Ownership for Ridesharing ($\beta_{\text{NoVehicle,Ridesharing}}$) & Ridesharing & All & 0.803 & 0.282 & 2.84 \\ 
No Vehicle Ownership for Transit ($\beta_{\text{NoVehicle,Transit}}$) & Transit & All & 0.74 & 0.22 & 3.37 \\ 
Usual Transit Users ($\beta_{\text{TransitUsual}}$) & Transit & All & 2.07 & 0.096 & 21.5 \\ 
Usual Ride-hailing User ($\beta_{\text{RidehailingUsual}}$) & Ridesharing & All & 0.905 & 0.251 & 3.61 \\ 
Cordon-based Pricing for Transit ($\beta_{\text{CordonPrice,Transit}}$) & Transit & All & 0.255 & 0.106 & 2.39 \\ 
Time-based Pricing for On-time Departure ($\beta_{\text{TimePrice,On-time}}$) & All modes & On-time & 0.327 & 0.078 & 4.19 \\ 
\midrule 
\multicolumn{3}{l}{Number of observations} & \multicolumn{3}{c}{4182} \\ 
\multicolumn{3}{l}{Number of parameters} & \multicolumn{3}{c}{35} \\ 
\multicolumn{3}{l}{Final log-likelihood} & \multicolumn{3}{c}{-6690.929} \\ 
\multicolumn{3}{l}{Rho-square} & \multicolumn{3}{c}{0.0915} \\ 
\multicolumn{3}{l}{Adjusted Rho-square} & \multicolumn{3}{c}{0.0867} \\ 
\multicolumn{3}{l}{AIC} & \multicolumn{3}{c}{13451.86} \\ 
\multicolumn{3}{l}{BIC} & \multicolumn{3}{c}{13673.67} \\ 
\botrule 
\end{tabular*} 
\end{sidewaystable}

\subsubsection{Nested logit model where modes are the nests}

To address MNL model limitations, an NL model was estimated, organizing alternatives into mode-based nests (Drive, Passenger, Ridesharing, Transit), with departure time options nested within each mode. The NL model results in Table \ref{tab:NL_results} shows consistent signs and significance levels compared to the MNL while also revealed improved behavioral realism. In particular, the estimated nesting parameters ($\mu$) for each mode were significantly different from unity, validating the existence of meaningful correlations among departure time alternatives within the same mode. Moreover, the magnitude of schedule delay penalties for early and late departures decreased compared to the MNL model, reflecting better accommodation of substitution across departure times within the same travel mode.

\begin{sidewaystable}
\caption{Nested Logit Model Estimation Results}\label{tab:NL_results}
\small
\renewcommand{\arraystretch}{1}
\begin{tabular*}{\textheight}{@{\extracolsep\fill}lllcccc}
\toprule
\textbf{Parameter Name} & \textbf{Mode} & \textbf{Dep. Time} & \textbf{Value} & \textbf{Std. Err.} & \textbf{t-test} & \textbf{p-value} \\
\midrule
Drive ASC & Driving & All & 1.250 & 0.205 & 6.120 & <0.001 \\
Ridesharing ASC & Ridesharing & All & -0.683 & 0.249 & -2.740 & 0.006 \\
Transit ASC & Transit & All & 0.218 & 0.266 & 0.817 & 0.414 \\
Scheduled Delay Cost for Early Departure & All modes & Early & -1.160 & 0.465 & -2.490 & 0.013 \\
Scheduled Delay Cost for Late Departure & All modes & Late & -2.050 & 0.820 & -2.500 & 0.012 \\
In Vehicle Travel Time for Driving & Driving & All & -0.023 & 0.011 & -2.070 & 0.039 \\
In Vehicle Travel Time for Passenger & Passenger & All & -0.028 & 0.011 & -2.560 & 0.011 \\
In Vehicle Travel Time for Ridesharing & Ridesharing & All & -0.035 & 0.012 & -2.970 & 0.003 \\
In Vehicle Time for Transit & Transit & All & -0.019 & 0.004 & -4.760 & <0.001 \\
Cost & All modes & All & -0.029 & 0.007 & -4.420 & <0.001 \\
Waiting Time & Ride, Pass., Transit & All & -0.005 & 0.002 & -2.460 & 0.014 \\
Toll (Congestion Pricing) & Drive, Passenger & All & -0.040 & 0.017 & -3.720 & <0.001 \\
Drive for Old Ages & Driving & All & 0.883 & 0.266 & 3.320 & 0.001 \\
Transit for Old Ages & Transit & All & 1.810 & 0.295 & 6.110 & <0.001 \\
Early Departure for Young Ages & All modes & Early & -0.256 & 0.121 & -2.120 & 0.034 \\
Male Passengers & Passenger & All & -0.365 & 0.088 & -4.140 & <0.001 \\
White Race for Driving & Driving & All & 0.258 & 0.075 & 3.420 & 0.001 \\
Snowy Weather for Early Departure & All modes & Early & 0.545 & 0.235 & 2.320 & 0.021 \\
Snowy Weather for Transit & Transit & All & -0.869 & 0.191 & -4.560 & <0.001 \\
Sunny Weather for Driving & Driving & All & 0.525 & 0.169 & 3.110 & 0.002 \\
Cold Weather for Transit & Transit & All & -0.895 & 0.193 & -4.640 & <0.001 \\
University Education for Driving & Driving & All & -0.406 & 0.076 & -5.370 & <0.001 \\
Free Parking for Driving & Driving & All & 0.384 & 0.100 & 3.840 & <0.001 \\
Household Size for Driving & Driving & All & -0.258 & 0.097 & -2.660 & 0.008 \\
Household Size for Passenger & Passenger & All & 0.289 & 0.108 & 2.670 & 0.008 \\
High Income for Late Departure & All modes & Late & 0.286 & 0.160 & 1.780 & 0.075 \\
High Income for Ridesharing & Ridesharing & All & -0.567 & 0.167 & -3.400 & 0.001 \\
Low Income for Driving & Driving & All & -0.458 & 0.158 & -2.900 & 0.004 \\
Vehicle Ownership for Driving & Driving & All & 0.599 & 0.079 & 7.610 & <0.001 \\
No Vehicle Ownership for Ridesharing & Ridesharing & All & 0.796 & 0.282 & 2.820 & 0.005 \\
No Vehicle Ownership for Transit & Transit & All & 0.721 & 0.220 & 3.280 & 0.001 \\
Usual Transit Users & Transit & All & 2.070 & 0.096 & 21.500 & <0.001 \\
Usual Ride-hailing User & Ridesharing & All & 0.919 & 0.251 & 3.660 & <0.001 \\
Cordon-based Pricing for Transit & Transit & All & 0.252 & 0.107 & 2.360 & 0.018 \\
Nesting Coefficient $\mu_{\text{Drive}}$ & Driving & All & 1.490 & 0.634 & 2.360 & 0.019 \\
Nesting Coefficient $\mu_{\text{Passenger}}$ & Passenger & All & 1.030 & 0.436 & 2.360 & 0.018 \\
Nesting Coefficient $\mu_{\text{Ridesharing}}$ & Ridesharing & All & 1.120 & 0.482 & 2.320 & 0.020 \\
Nesting Coefficient $\mu_{\text{Transit}}$ & Transit & All & 1.490 & 0.617 & 2.420 & 0.015 \\
\midrule
\multicolumn{3}{l}{Number of observations}   & \multicolumn{4}{c}{4182} \\
\multicolumn{3}{l}{Number of parameters}     & \multicolumn{4}{c}{38} \\
\multicolumn{3}{l}{Final log-likelihood}     & \multicolumn{4}{c}{-6680.010} \\
\multicolumn{3}{l}{Rho-square}               & \multicolumn{4}{c}{0.0899} \\
\multicolumn{3}{l}{Adjusted Rho-square}      & \multicolumn{4}{c}{0.0847} \\
\multicolumn{3}{l}{AIC}                      & \multicolumn{4}{c}{13436.02} \\
\multicolumn{3}{l}{BIC}                      & \multicolumn{4}{c}{13676.88} \\
\botrule
\end{tabular*}
\end{sidewaystable}

Both models confirmed that travel cost, tolls, travel time, and waiting time have negative impacts on utility, while habitual transit and ridesharing usage positively influence transit and rideshare mode choice respectively. Pricing strategies such as cordon-based also showed significant effects on selecting transit mode, supporting the potential effectiveness of congestion pricing in influencing travel behavior. Significant nest scale parameters confirm the appropriateness of the nesting structure. Compared to the MNL, the NL model produces slightly attenuated estimates for schedule delay penalties and better reflects travellers' substitution patterns, offering a more behaviourally nuanced model of travel decisions under congestion pricing scenarios.

\subsubsection{Cross-nested logit model where modes and departure times are the nests}

The CNL model was developed to account for the fact that travel alternatives can simultaneously belong to both mode and departure time nests. As shown in Table 5, the estimated cross-nesting parameters indicate travelers consider both mode and timing when making decisions, which underscores the importance of capturing multidimensional substitution patterns of mode and departure time.

The comparative analysis of the MNL, NL, and CNL models reveals that model performance improves as the specification allows for more flexible substitution patterns among alternatives.  The MNL model, serving as the baseline, achieves a final log-likelihood of –6694.36. Incorporating nesting structures, the NL model achieves a slightly better fit (–6680.01), while the CNL model further improves the log-likelihood to –6650.92. This progressive improvement in model fit is supported by goodness-of-fit indicators: the CNL model yields the lowest Akaike Information Criterion (AIC = 13,407.84) compared to the NL (13,436.02) and MNL (13,460.72) models, indicating a better trade-off between fit and complexity. 

\begin{sidewaystable}
\caption{Cross-Nested Logit Model Estimation Results}\label{tab:CNL_results}
\small
\renewcommand{\arraystretch}{0.9}
\begin{tabular*}{\textheight}{@{\extracolsep\fill}lllcccc}
\toprule
\textbf{Parameter Name} & \textbf{Mode} & \textbf{Dep. Time} &
\textbf{Value} & \textbf{Std. Err.} & \textbf{t-test} & \textbf{p-value} \\
\midrule
Drive ASC & Driving & All & 0.844 & 0.204 & 4.140 & <0.001 \\
Ridesharing ASC & Ridesharing & All & -0.940 & 0.541 & -1.740 & 0.082 \\
Transit ASC & Transit & All & -0.183 & 0.270 & -0.676 & 0.499 \\
Scheduled Delay Cost for Early Departure & All modes & Early & -1.180 & 0.298 & -3.950 & <0.001 \\
Scheduled Delay Cost for Late Departure & All modes & Late & -2.080 & 0.463 & -4.490 & <0.001 \\
In Vehicle Travel Time for Driving & Driving & All & -0.022 & 0.011 & -2.060 & 0.039 \\
In Vehicle Travel Time for Passenger & Passenger & All & -0.028 & 0.011 & -2.630 & 0.008 \\
In Vehicle Travel Time for Ridesharing & Ridesharing & All & -0.036 & 0.011 & -3.170 & 0.002 \\
In Vehicle Time for Transit & Transit & All & -0.020 & 0.004 & -5.490 & <0.001 \\
Cost & All modes & All & -0.028 & 0.007 & -4.320 & <0.001 \\
Waiting Time & Ride, Pass., Transit & All & -0.004 & 0.002 & -2.340 & 0.019 \\
Toll (Congestion Pricing) & Drive, Passenger & All & -0.040 & 0.011 & -3.590 & <0.001 \\
Drive for Old Ages & Driving & All & 0.869 & 0.265 & 3.280 & 0.001 \\
Transit for Old Ages & Transit & All & 1.790 & 0.293 & 6.100 & <0.001 \\
Early Departure for Young Ages & All modes & Early & -0.195 & 0.090 & -2.160 & 0.031 \\
Male Passengers & Passenger & All & -0.350 & 0.088 & -3.970 & <0.001 \\
White Race for Driving & Driving & All & 0.262 & 0.074 & 3.540 & <0.001 \\
Snowy Weather for Early Departure & All modes & Early & 0.480 & 0.122 & 3.940 & <0.001 \\
Snowy Weather for Transit & Transit & All & -0.895 & 0.189 & -4.750 & <0.001 \\
Sunny Weather for Driving & Driving & All & 0.538 & 0.169 & 3.190 & 0.001 \\
Cold Weather for Transit & Transit & All & -0.904 & 0.192 & -4.720 & <0.001 \\
University Education for Driving & Driving & All & -0.391 & 0.075 & -5.240 & <0.001 \\
Free Parking for Driving & Driving & All & 0.371 & 0.099 & 3.750 & <0.001 \\
Household Size for Driving & Driving & All & -0.258 & 0.095 & -2.720 & 0.006 \\
Household Size for Passenger & Passenger & All & 0.271 & 0.107 & 2.540 & 0.011 \\
High Income for Late Departure & All modes & Late & 0.347 & 0.131 & 2.650 & 0.008 \\
High Income for Ridesharing & Ridesharing & All & -0.579 & 0.169 & -3.430 & 0.001 \\
Low Income for Driving & Driving & All & -0.410 & 0.159 & -2.580 & 0.010 \\
Vehicle Ownership for Driving & Driving & All & 0.575 & 0.079 & 7.300 & <0.001 \\
No Vehicle Ownership for Ridesharing & Ridesharing & All & 0.951 & 0.290 & 3.280 & 0.001 \\
No Vehicle Ownership for Transit & Transit & All & 0.850 & 0.227 & 3.740 & <0.001 \\
Usual Transit Users & Transit & All & 2.030 & 0.100 & 20.300 & <0.001 \\
Usual Ride-hailing User & Ridesharing & All & 0.915 & 0.251 & 3.640 & <0.001 \\
Cordon-based Pricing for Transit & Transit & All & 0.254 & 0.103 & 2.460 & 0.014 \\
Nesting Coefficient $\mu_{\text{Drive}}$ & Driving & All & 9.290 & 4.210 & 2.210 & 0.027 \\
Nesting Coefficient $\mu_{\text{Passenger}}$ & Passenger & All & 1.560 & 0.469 & 3.320 & 0.001 \\
Nesting Coefficient $\mu_{\text{Ridesharing}}$ & Ridesharing & All & 1.480 & 0.616 & 2.400 & 0.017 \\
Nesting Coefficient $\mu_{\text{Transit}}$ & Transit & All & 4.160 & 4.670 & 0.891 & 0.373 \\
Nesting Coefficient $\mu_{\text{Early}}$ & All & Early & 1.190 & 0.128 & 9.290 & <0.001 \\
Nesting Coefficient $\mu_{\text{On-time}}$ & All & On-time & 4.830 & 5.250 & 0.920 & 0.357 \\
Nesting Coefficient $\mu_{\text{Late}}$ & All & Late & 1.000 & 0.133 & 7.540 & <0.001 \\
\midrule
\multicolumn{3}{l}{Number of observations}   & \multicolumn{4}{c}{4182} \\
\multicolumn{3}{l}{Number of parameters}     & \multicolumn{4}{c}{53} \\
\multicolumn{3}{l}{Final log-likelihood}     & \multicolumn{4}{c}{-6650.920} \\
\multicolumn{3}{l}{Rho-square}               & \multicolumn{4}{c}{0.0900} \\
\multicolumn{3}{l}{Adjusted Rho-square}      & \multicolumn{4}{c}{0.0828} \\
\multicolumn{3}{l}{AIC}                      & \multicolumn{4}{c}{13407.84} \\
\multicolumn{3}{l}{BIC}                      & \multicolumn{4}{c}{13743.78} \\
\botrule
\end{tabular*}
\end{sidewaystable}

Overall, the progression from MNL to CNL demonstrated improved behavioral realism, particularly by relaxing the restrictive IIA assumption and better captures substitution between travel alternatives. Ultimately, model fit, with the CNL model offering a comprehensive and flexible representation of the joint mode and departure time decision process under congestion pricing conditions. The majority of estimated coefficients are statistically significant at conventional levels (95\% confidence), and the signs align with behavioral expectations, enhancing the credibility of the model.

Travel cost, tolls, in-vehicle travel times and waiting time have negative and significant coefficients, confirming their negative influence on the utility of alternatives, as expected. Departure time preferences are also evident. The coefficients for scheduled delay penalties are negative and statistically significant: $\beta_{\text{Early}}=-1.180$ and $\beta_{\text{Late}}=-2.080$. The larger magnitude for late departures indicates that travelers have a stronger disutility for being late compared to being early, a result consistent with standard scheduling models in transportation behavior \citep{thorhauge2016flexible}.

Socio-demographic characteristics significantly affect choice behavior. Older respondents (Age 65+) have a positive association with both driving and transit modes, while individuals with university education are less likely to choose driving (-0.391). Males are less likely to choose passenger travel (-0.350). Household size influences choices differently across modes: larger households decrease the utility of driving but increase the utility of choosing passenger modes.

Income effects are observed: higher-income individuals are more likely to delay their departure (0.347) but less likely to choose ridesharing (-0.579). Low-income individuals show a lower propensity to drive (-0.410), possibly reflecting financial constraints associated with car use. Weather conditions significantly influence choices. Snowy weather increases the likelihood of early departures (0.480) but reduces the utility of transit travel (-0.895). Sunny weather positively affects the choice of driving (0.538), suggesting that favourable weather encourages private vehicle use. Behavioral inertia is captured through significant positive coefficients for habitual users: those who usually use transit (2.030) and those who usually use ridesharing (0.915) are more likely to continue using their familiar modes. Pricing policy variables are significant. Cordon-based pricing positively influences transit use (0.254), indicating that congestion pricing policies could shift demand towards transit. 

The nest scale parameters ($\mu$) from the CNL model indicate varying degrees of similarity and substitutability among alternatives within each nest. Specifically, the high scale parameters observed for driving ($\mu = 9.290$), on-time departure ($\mu = 4.830$), and transit ($\mu = 4.160$) imply low variance within these nests, reflecting high correlation and substantial substitutability among alternatives in the same nest. Travelers thus perceive options within these nests as more similar or interchangeable. Conversely, nests with lower scale parameters, such as passenger ($\mu = 1.560$), ridesharing ($\mu = 1.480$), and early departure ($\mu = 1.190$), exhibit higher variance and consequently lower correlation, indicating greater independence and differentiation between alternatives. The late departure nest, with $\mu$ fixed at 1.0, behaves similarly to the standard MNL model, signifying complete independence of alternatives within that nest.
 
In the CNL model, the $\alpha$ parameters reflect the degree to which each alternative shares unobserved characteristics with others in a given nest, either mode-based or departure-time-based as illustrated in Table \ref{tab:alpha_membership}. A higher $\alpha$ value for a particular nest indicates that the alternative is more strongly associated with the behavioural dimension that the nest represents. Consequently, a “greater likelihood of substitution within that nest” implies that if an alternative becomes less attractive (due to, for example, increased cost or travel time), travellers are more likely to switch to another alternative within the same nest rather than outside of it. For instance, if an alternative has a high $\alpha$ value in the mode nest, it suggests that mode plays a dominant role in choice behaviour, and substitution will occur among alternatives of the same mode rather than across different modes. Conversely, a high $\alpha$ in the departure-time nest indicates that timing is the more influential factor, and travellers are more likely to switch between early, on-time, or late options regardless of mode. This cross-nested structure captures the nuanced interplay between mode and timing in traveller decision-making under congestion pricing scenarios.

\begin{table*}[h]
\caption{Cross-Nested Membership Parameters and Interpretation}
\label{tab:alpha_membership}
\footnotesize
\renewcommand{\arraystretch}{1.2}

\begin{tabular*}{\textwidth}{@{\extracolsep\fill}
p{0.05\textwidth}   % Alt
p{0.1\textwidth}   % Mode
p{0.09\textwidth}   % Dep time
c                   % alpha Mode
c                   % alpha Time
c                   % p-value
p{0.44\textwidth}   % Interpretation
}
\toprule
\textbf{Alt.} & 
\textbf{Mode Nest} &
\textbf{Dep. Time Nest} &
$\boldsymbol{\alpha_{\text{Mode}}}$ &
$\boldsymbol{\alpha_{\text{Time}}}$ &
\textbf{P-value} &
\textbf{Interpretation} \\
\midrule

V1  & Drive       & Early     & 0.452 & 0.548 & 0.006 & Shared membership indicates substitution flexibility across driving and early departure; travellers perceive other early options as comparable. \\

V2  & Drive       & On-time   & 0.237 & 0.763 & 0.166 & Weak Drive-mode association; substitution appears driven more by timing than by mode. \\

V3  & Drive       & Late      & 0.580 & 0.419 & 0.002 & Stronger identification with Drive mode; limited substitution across late departures. \\

V4  & Passenger   & Early     & 0.916 & 0.084 & 0.001 & Very strong Passenger-mode identity; early-departure alternatives are not perceived as substitutes. \\

V5  & Passenger   & On-time   & 0.757 & 0.243 & 0.000 & High Passenger-mode membership with moderate on-time departure flexibility. \\

V6  & Passenger   & Late      & 0.219 & 0.781 & 0.622 & Weak mode association; substitution is primarily across time rather than across modes. \\

V7  & Ridesharing & Early     & 0.861 & 0.139 & 0.113 & Moderate ridesharing identity with limited substitution among early options. \\

V8  & Ridesharing & On-time   & 1.000 & 0.000 & 0.044 & Fully nested in mode; travellers do not substitute based on timing. \\

V9  & Ridesharing & Late      & 0.000 & 1.000 & --    & No mode-based membership; substitution is strictly along the time dimension. \\

V10 & Transit     & Early     & 0.133 & 0.867 & 0.620 & Weak Transit-mode substitution; early departure time dominates choice behaviour. \\

V11 & Transit     & On-time   & 0.995 & 0.005 & 0.000 & Very strong mode identity; viewed as a distinct alternative. \\

V12 & Transit     & Late      & 0.786 & 0.214 & 0.000 & Strong Transit-mode membership with moderate timing flexibility. \\

\botrule
\end{tabular*}

\end{table*}

The $\alpha$ parameter estimates presented in the table reveal important insights into the behavioural structure of substitution patterns across both mode and departure-time dimensions. High $\alpha$ values such as for Passenger–Early (0.916) and Transit–On time (0.995) indicate strong associations with their respective mode nests, suggesting that users selecting these alternatives are more likely to consider other options within the same mode rather than switch across modes. In contrast, alternatives like Transit–Early (0.133) or Ridesharing–Late (0.000)(this alternative is entirely nested under a single branch) demonstrate weak mode membership and instead derive most of their nesting weight from the departure-time dimension. This implies that for these users, departure timing is a more salient factor than mode, and they are more likely to substitute between modes as long as the departure time remains the same. The variation in $\alpha$ values across alternatives highlights how travelers weigh mode and timing differently depending on the context of the trip. These findings support the need for congestion pricing policies that simultaneously account for both temporal and modal dimensions. For example, a hybrid pricing scheme could involve higher toll rates specifically during peak commuting periods to incentivize temporal shifts, coupled with reduced fares or dedicated incentives for transit and ridesharing options to encourage mode shifts among travelers less responsive to time-based pricing alone. This dual approach acknowledges the behavioral heterogeneity revealed in traveller responses, maximizing overall policy effectiveness.

The consistency of parameter estimates between the NL and CNL models indicates that the fundamental determinants of travel and departure time choices are robust across different model structures. This stability reinforces the reliability of the model specification and the validity of the inferred policy insights. Several $\alpha$ parameters in the CNL model, while not statistically significant at conventional levels (e.g., $\alpha_{\text{Drive}}^{\text{On time}}=0.237$; $\alpha_{\text{Passenger}}^{\text{Late}}=0.219$), were retained in the model due to their theoretical relevance and contribution to behavioural interpretation. As emphasized by Hess et al. (2025), statistical insignificance should not automatically warrant exclusion; parameters with theoretically consistent signs and plausible magnitudes can still convey important behavioural patterns. Moreover, significance levels are sensitive to sample size, collinearity, and model specification. In this context, the inclusion of such α parameters supports a richer representation of overlapping substitution patterns across mode and departure time nests, which is central to the flexibility offered by the CNL framework.

Compared to the NL model, the CNL model offers a more flexible and behaviourally nuanced representation of travel choices under congestion pricing. While core behavioural drivers such as travel cost, time, tolls, socio-demographic attributes, and weather conditions remain consistent across both models, the CNL model introduces additional substitution patterns across both mode and departure time dimensions. The magnitude of the alternative-specific constants adjusts slightly under the CNL structure, reflecting a more detailed representation of choice behaviour. Moreover, the estimated cross-nesting parameters and significant scale parameters in the CNL model confirm that travellers simultaneously consider both how and when to travel when making their decisions. The CNL model thus captures a richer substitution structure compared to the NL model, providing a superior behavioural representation for policy analysis and planning.

\subsubsection{Policy Implications for Congestion Pricing}

The results of the CNL model, supported by the MNL model findings, highlight the multifaceted behavioural responses to congestion pricing. Toll increases and higher driving costs significantly reduce the attractiveness of driving, particularly for lower-income individuals, prompting shifts toward more affordable options such as transit and ridesharing. Peak-period pricing encourages some travellers to adjust their schedules, with a clear preference for earlier departures to avoid late penalties, reflecting the strong disutility associated with arriving late and the comparatively milder aversion to being early. Notably, the MNL model shows that time-based pricing applied specifically to on-time departures has a significant positive effect on behavioural change, indicating that even modest tolls during the peak can incentivize time shifts, particularly among those with flexible schedules. Cordon pricing, meanwhile, is associated with increased transit usage, underscoring its effectiveness in redirecting demand away from private vehicles and into more sustainable modes. Environmental and habitual factors also modulate responses, sunny conditions slightly favour driving, while snowy weather increases the likelihood of early departures and enhances substitution across modes. Together, these results affirm the potential of well-calibrated congestion pricing, especially when combining cordon-based and time-targeted approaches, to reduce peak congestion through both modal shifts and departure time adjustments, while maintaining equity and behavioural realism. Three primary pricing schemes were considered: cordon-based pricing, distance-based pricing, and travel-time-based pricing.

The significant positive coefficient for cordon-based pricing for a significant  mode shift towards transit confirms that implementing a cordon charge around high-congestion zones, such as city centres, effectively shifts demand away from solo driving. Given the observed strong mode commitment among certain travellers, particularly Transit On-time and Passenger Early, cordon pricing appears particularly effective in reducing private vehicle use within targeted areas. Policymakers can leverage this finding by strategically placing cordon tolling around heavily congested urban cores, thereby promoting transit use, alleviating downtown traffic, and enhancing overall urban sustainability.

The CNL model did not yield significant results for distance-based pricing variables, suggesting limited behavioural responsiveness when tolls are structured strictly by travel distance   . This lack of significance indicates that travellers might perceive distance-based tolls as less transparent or salient, reducing their effectiveness as behaviour-change tools. Therefore, distance-based tolling alone may have limited utility for congestion management in urban environments unless combined with more targeted, transparent, and psychologically impactful approaches, such as cordon or peak-period pricing.

Interestingly, while the CNL model did not show significant results for travel time-based pricing, the MNL model indicated a significant positive impact for on-time departures. This suggests that travellers, especially those commuting during typical peak periods, may still be responsive to temporal toll adjustments by shifting their departure time. Even though this effect diminishes when accounting for complex substitution patterns in the CNL framework, it still implies potential utility for peak-hour tolling schemes aimed at shifting travellers to off-peak periods. Policymakers should cautiously consider time-based pricing, ensuring clarity, simplicity, and effective communication to enhance traveller responsiveness.

The estimated $\mu$ and $\alpha$ parameters offer nuanced insights into traveller substitution patterns. High $\mu$ values for Drive and On-time departures indicate limited flexibility in altering choices within these nests. Travellers who prefer driving and departing at typical peak times display strong behavioural inertia, thus targeted incentives, such as significant price differentiation between peak and off-peak periods, may be necessary to stimulate behavioural shifts. 
In contrast, lower $\mu$ values for nests like Late and Early suggest greater flexibility, indicating that travellers may more easily substitute between different modes or time alternatives within these nests. Therefore, policymakers could effectively use off-peak discounts or early/late travel incentives to redistribute travel demand away from peak periods.

Alpha parameters reinforce the importance of considering multidimensional substitution. For example, strong mode-based substitution (high $\alpha$ values) such as in Passenger Early and Transit On-time implies that mode-specific policies such as transit service enhancements, fare adjustments, or targeted subsidies are essential to support these loyal user groups. On the other hand, alternatives like Transit Early or Ridesharing Late reflect strong temporal substitution potential, underscoring the value of time-based incentives to influence these travellers effectively.

Model estimates further highlight that congestion pricing impacts vary by socio-demographic characteristics and environmental conditions. Lower-income travellers show strong sensitivity to increased costs and tolls, suggesting a need for equity-oriented measures, such as discounts or exemptions targeted at lower-income commuters, to ensure that congestion pricing schemes do not disproportionately burden these groups.Furthermore, weather conditions significantly influence traveller responses; snowy conditions notably enhance early departures, while sunny conditions increase the attractiveness of driving. Congestion pricing schemes could thus incorporate weather-responsive pricing or information systems to dynamically adjust policies based on real-time conditions.

Based on the CNL model’s detailed findings, a hybrid congestion pricing strategy appears most effective. This approach should combine cordon-based pricing to directly target central congestion zones and encourage transit use with peak-period time-based pricing incentives to shift traveller departure times, particularly given the rigid preferences observed during peak commuting windows. Additionally, complementary measures like improved public transport services, equity-based discounts, and real-time traveller information systems could enhance overall effectiveness, fairness, and public acceptability of the policy.

\subsection{Elasticity Analysis}

\subsubsection{Analysis of Time Elasticities}

The elasticity results presented in  Tables \ref{tab:time_elasticity_mnl_cnl} capture travellers' responsiveness to changes in travel times for various alternatives. Elasticities quantify the percent change in the probability of choosing a specific alternative in response to a one-percent change in the travel time of the same (direct elasticity) or another alternative (cross elasticity). Negative values indicate a decrease in attractiveness with increased travel time, while positive values indicate substitution to other alternatives. Highlighted green cells in Table \ref{tab:time_elasticity_mnl_cnl}.b indicate those time elasticities that exhibit a higher magnitude in the CNL model compared to the MNL model, illustrating the enhanced sensitivity captured by the CNL specification.

Among the driving alternatives, late departure has the highest direct elasticity magnitude, followed by early and on-time departure. This suggests that drivers who choose to depart early or late exhibit a stronger behavioural response to changes in travel time, likely due to tighter schedule constraints or greater flexibility, compared to those who depart on time, who may be more committed to fixed peak-hour routines and thus less reactive to marginal travel time increases. \textit{Passenger}$_\text{Early}$ and \textit{Passenger}$_\text{On time}$ also show substantial sensitivity, suggesting passengers strongly prefer predictable, shorter travel times. For ridesharing, \textit{Ridesharing}$_\text{Early}$ and \textit{Ridesharing}$_\text{On time}$ display high sensitivity to changes in travel time, highlighting users’ aversion to longer durations. Transit users are similarly affected, with significant negative elasticities indicating that increases in transit travel time substantially reduce their likelihood of choosing these options.

\begin{sidewaystable}
\caption{Time elasticity estimates for the MNL and CNL models}
\label{tab:time_elasticity_mnl_cnl}
\footnotesize
\renewcommand{\arraystretch}{1}

% ==================== (a) MNL ====================
\begin{tabular*}{\textheight}{@{\extracolsep\fill}lrrrrrrrrrrrr}
\toprule
\multicolumn{13}{l}{\textbf{(a) Time Elasticity for the MNL Model}}\\[4pt]
\textbf{Alternative / Variable} &
\shortstack{Drive\\Early} &
\shortstack{Drive\\On-time} &
\shortstack{Drive\\Late} &
\shortstack{Pass\\Early} &
\shortstack{Pass\\On-time} &
\shortstack{Pass\\Late} &
\shortstack{Ride\\Early} &
\shortstack{Ride\\On-time} &
\shortstack{Ride\\Late} &
\shortstack{Transit\\Early} &
\shortstack{Transit\\On-time} &
\shortstack{Transit\\Late} \\
\midrule
Drive-Early        & -0.360 & 0.213 & 0.017 & 0.021 & 0.090 & 0.006 & 0.008 & 0.029 & 0.002 & 0.018 & 0.068 & 0.005 \\
Drive-On-time      & 0.052  & -0.238 & 0.017 & 0.018 & 0.078 & 0.006 & 0.007 & 0.029 & 0.002 & 0.016 & 0.071 & 0.005 \\
Drive-Late         & 0.054  & 0.225 & -0.379 & 0.019 & 0.083 & 0.007 & 0.007 & 0.029 & 0.002 & 0.016 & 0.072 & 0.005 \\
Passenger-Early    & 0.053  & 0.191 & 0.015 & -0.566 & 0.096 & 0.007 & 0.010 & 0.037 & 0.003 & 0.022 & 0.085 & 0.006 \\
Passenger-On-time  & 0.046  & 0.194 & 0.015 & 0.023 & -0.543 & 0.008 & 0.009 & 0.037 & 0.003 & 0.020 & 0.089 & 0.006 \\
Passenger-Late     & 0.048  & 0.202 & 0.017 & 0.024 & 0.102 & -0.551 & 0.009 & 0.036 & 0.003 & 0.020 & 0.090 & 0.007 \\
Ridesharing-Early  & 0.047  & 0.166 & 0.012 & 0.023 & 0.083 & 0.006 & -0.742 & 0.042 & 0.003 & 0.025 & 0.091 & 0.006 \\
Ridesharing-On-time& 0.041  & 0.167 & 0.013 & 0.020 & 0.085 & 0.006 & 0.010 & -0.771 & 0.003 & 0.022 & 0.094 & 0.006 \\
Ridesharing-Late   & 0.043  & 0.175 & 0.014 & 0.021 & 0.089 & 0.007 & 0.010 & 0.042 & -0.705 & 0.022 & 0.095 & 0.006 \\
Transit-Early      & 0.032  & 0.120 & 0.009 & 0.016 & 0.061 & 0.005 & 0.008 & 0.030 & 0.002 & -0.536 & 0.194 & 0.013 \\
Transit-On-time    & 0.028  & 0.121 & 0.009 & 0.014 & 0.063 & 0.005 & 0.007 & 0.029 & 0.002 & 0.044 & -0.423 & 0.014 \\
Transit-Late       & 0.029  & 0.127 & 0.010 & 0.015 & 0.065 & 0.005 & 0.007 & 0.029 & 0.002 & 0.045 & 0.204 & -0.521 \\
\botrule
\end{tabular*}

\vspace{1cm}

% ==================== (b) CNL ====================
\begin{tabular*}{\textheight}{@{\extracolsep\fill}lrrrrrrrrrrrr}
\toprule
\multicolumn{13}{l}{\textbf{(b) Time Elasticity for the CNL Model}}\\[4pt]
\textbf{Alternative / Variable} &
\shortstack{Drive\\Early} &
\shortstack{Drive\\On-time} &
\shortstack{Drive\\Late} &
\shortstack{Pass\\Early} &
\shortstack{Pass\\On-time} &
\shortstack{Pass\\Late} &
\shortstack{Ride\\Early} &
\shortstack{Ride\\On-time} &
\shortstack{Ride\\Late} &
\shortstack{Transit\\Early} &
\shortstack{Transit\\On-time} &
\shortstack{Transit\\Late} \\
\midrule
Drive-Early &
\cellcolor{green!20}-0.458 & \cellcolor{green!20}0.302 & \cellcolor{green!20}0.020 &
0.019 & 0.065 & \cellcolor{green!20}0.009 &
0.006 & 0.020 & \cellcolor{green!20}0.003 &
0.018 & 0.055 & 0.002 \\
Drive-On-time &
\cellcolor{green!20}0.067 & \cellcolor{green!20}-0.279 & \cellcolor{green!20}0.026 &
0.014 & 0.057 & \cellcolor{green!20}0.009 &
0.004 & 0.019 & \cellcolor{green!20}0.003 &
0.016 & 0.063 & 0.002 \\
Drive-Late &
\cellcolor{green!20}0.071 & \cellcolor{green!20}0.419 & \cellcolor{green!20}-0.567 &
0.015 & 0.059 & \cellcolor{green!20}0.010 &
0.004 & 0.019 & \cellcolor{green!20}0.003 &
0.016 & 0.059 & 0.002 \\
Passenger-Early &
\cellcolor{green!20}0.064 & \cellcolor{green!20}0.205 & 0.014 &
-0.450 & 0.070 & \cellcolor{green!20}0.011 &
0.007 & 0.025 & \cellcolor{green!20}0.004 &
0.022 & 0.072 & 0.002 \\
Passenger-On-time &
0.046 & \cellcolor{green!20}0.216 & 0.014 &
0.018 & -0.455 & \cellcolor{green!20}0.011 &
0.005 & 0.025 & \cellcolor{green!20}0.004 &
0.020 & \cellcolor{green!20}0.116 & 0.002 \\
Passenger-Late &
0.047 & \cellcolor{green!20}0.216 & 0.015 &
0.018 & 0.074 & -0.417 &
0.005 & 0.024 & \cellcolor{green!20}0.004 &
0.020 & 0.075 & 0.002 \\
Ridesharing-Early &
\cellcolor{green!20}0.065 & \cellcolor{green!20}0.181 & 0.011 &
0.022 & 0.062 & \cellcolor{green!20}0.009 &
-0.534 & 0.029 & \cellcolor{green!20}0.004 &
0.025 & 0.077 & 0.002 \\
Ridesharing-On-time &
\cellcolor{green!20}0.042 & \cellcolor{green!20}0.185 & 0.011 &
0.016 & 0.063 & \cellcolor{green!20}0.009 &
0.006 & -0.529 & \cellcolor{green!20}0.004 &
0.022 & 0.079 & 0.002 \\
Ridesharing-Late &
0.043 & \cellcolor{green!20}0.191 & 0.012 &
0.017 & 0.065 & \cellcolor{green!20}0.010 &
0.006 & 0.028 & -0.477 &
0.022 & 0.079 & 0.002 \\
Transit-Early &
0.032 & \cellcolor{green!20}0.131 & 0.008 &
0.012 & 0.044 & \cellcolor{green!20}0.007 &
0.004 & 0.020 & \cellcolor{green!20}0.003 &
-0.448 & 0.159 & \cellcolor{green!20}0.014 \\
Transit-On-time &
0.027 & \cellcolor{green!20}0.141 & 0.008 &
0.011 & 0.061 & \cellcolor{green!20}0.007 &
0.004 & 0.020 & \cellcolor{green!20}0.003 &
0.043 & -0.389 & 0.005 \\
Transit-Late &
0.029 & \cellcolor{green!20}0.140 & 0.009 &
0.012 & 0.047 & \cellcolor{green!20}0.008 &
0.004 & 0.019 & \cellcolor{green!20}0.003 &
\cellcolor{green!20}0.146 & 0.167 & -0.519 \\
\botrule
\end{tabular*}

\end{sidewaystable}

Notably, driving alternatives exhibit substantial cross-elasticity within the driving mode, suggesting that drivers switch between different departure times when faced with increased travel times. Moderate cross elasticities exist between modes, such as driving and passenger or driving and transit, indicating some potential to shift between these modes under changing time conditions. However, intra-mode (within same mode in different departure times) substitution is more dominant.

Higher direct elasticity in the CNL model (e.g., \textit{Drive}$_\text{Late}$: $-0.5672$ in CNL vs. $-0.3792$ in MNL) indicates better sensitivity capture, showing travellers' stronger behavioural response to time increases. Cross-elasticities in the CNL model are generally higher and more realistic, indicating better representation of substitution effects and capturing travellers' realistic behaviour of shifting between alternatives when faced with higher travel times.

\subsubsection{Analysis of Cost Elasticities}

Based on the results shown in Tables \ref{tab:cost_elasticity_mnl_cnl}, elasticities for driving alternatives indicate moderate sensitivity to driving costs, especially during on-time and late departures. Highlighted green cells in Table \ref{tab:cost_elasticity_mnl_cnl}.b represent those cost elasticities that exhibit greater magnitude in the CNL model than in the MNL model. Early departures show less elasticity, suggesting travellers departing early are somewhat price-sensitive but less so than late departures. Cross-elasticities within driving alternatives indicate that increased driving costs during peak periods (on-time) could effectively shift drivers toward earlier or later departure times, helping alleviate peak congestion.

Passenger alternatives show moderate responsiveness to their own costs, with on-time departures having relatively high elasticity. Significant cross-elasticities between passenger and driving indicate that raising passenger costs could inadvertently shift users back toward driving, highlighting the importance of keeping passenger mode pricing competitive and attractive.
Ridesharing elasticities are higher compared to driving and passenger modes, implying that ridesharing users are cost-sensitive and may quickly shift modes or departure times in response to increased pricing. Cross-elasticities indicate that increased ridesharing costs mildly push travellers towards passenger or driving alternatives.

\begin{sidewaystable}
\caption{Cost elasticity estimates for the MNL and CNL models}
\label{tab:cost_elasticity_mnl_cnl}
\footnotesize
\renewcommand{\arraystretch}{1}

% ==================== (a) MNL ====================
\begin{tabular*}{\textheight}{@{\extracolsep\fill}lrrrrrrrrrrrr}
\toprule
\multicolumn{13}{l}{\textbf{(a) Cost Elasticity for the MNL Model}}\\[4pt]
\textbf{Alternative / Variable} &
\shortstack{Drive\\Early} &
\shortstack{Drive\\On-time} &
\shortstack{Drive\\Late} &
\shortstack{Pass\\Early} &
\shortstack{Pass\\On-time} &
\shortstack{Pass\\Late} &
\shortstack{Ride\\Early} &
\shortstack{Ride\\On-time} &
\shortstack{Ride\\Late} &
\shortstack{Transit\\Early} &
\shortstack{Transit\\On-time} &
\shortstack{Transit\\Late} \\
\midrule
Drive-Early       & -0.203 & 0.111 & 0.011 & 0.005 & 0.015 & 0.001 & 0.003 & 0.009 & 0.001 & 0.003 & 0.009 & 0.001 \\
Drive-On-time     &  0.029 & -0.120 & 0.011 & 0.004 & 0.015 & 0.001 & 0.002 & 0.009 & 0.001 & 0.002 & 0.010 & 0.001 \\
Drive-Late        &  0.032 & 0.118 & -0.234 & 0.004 & 0.016 & 0.002 & 0.002 & 0.009 & 0.001 & 0.002 & 0.009 & 0.001 \\
Passenger-Early   &  0.031 & 0.100 & 0.010 & -0.116 & 0.018 & 0.002 & 0.004 & 0.012 & 0.001 & 0.003 & 0.011 & 0.001 \\
Passenger-On-time &  0.027 & 0.100 & 0.010 & 0.005 & -0.100 & 0.002 & 0.003 & 0.011 & 0.001 & 0.003 & 0.011 & 0.001 \\
Passenger-Late    &  0.028 & 0.106 & 0.011 & 0.005 & 0.020 & -0.123 & 0.003 & 0.011 & 0.001 & 0.003 & 0.011 & 0.001 \\
Ridesharing-Early &  0.027 & 0.084 & 0.008 & 0.005 & 0.015 & 0.001 & -0.240 & 0.013 & 0.001 & 0.004 & 0.012 & 0.001 \\
Ridesharing-On-time &
                    0.023 & 0.084 & 0.008 & 0.004 & 0.015 & 0.001 & 0.003 & -0.234 & 0.001 & 0.003 & 0.013 & 0.001 \\
Ridesharing-Late  &  0.024 & 0.088 & 0.008 & 0.004 & 0.016 & 0.002 & 0.003 & 0.013 & -0.239 & 0.003 & 0.013 & 0.001 \\
Transit-Early     &  0.017 & 0.057 & 0.005 & 0.003 & 0.011 & 0.001 & 0.003 & 0.009 & 0.001 & -0.075 & 0.024 & 0.002 \\
Transit-On-time   &  0.014 & 0.056 & 0.005 & 0.003 & 0.011 & 0.001 & 0.002 & 0.008 & 0.001 & 0.006 & -0.055 & 0.002 \\
Transit-Late      &  0.015 & 0.060 & 0.006 & 0.003 & 0.011 & 0.001 & 0.002 & 0.008 & 0.001 & 0.006 & 0.024 & -0.079 \\
\botrule
\end{tabular*}

\vspace{1cm}

% ==================== (b) CNL ====================
\begin{tabular*}{\textheight}{@{\extracolsep\fill}lrrrrrrrrrrrr}
\toprule
\multicolumn{13}{l}{\textbf{(b) Cost Elasticity for the CNL Model}}\\[4pt]
\textbf{Alternative / Variable} &
\shortstack{Drive\\Early} &
\shortstack{Drive\\On-time} &
\shortstack{Drive\\Late} &
\shortstack{Pass\\Early} &
\shortstack{Pass\\On-time} &
\shortstack{Pass\\Late} &
\shortstack{Ride\\Early} &
\shortstack{Ride\\On-time} &
\shortstack{Ride\\Late} &
\shortstack{Transit\\Early} &
\shortstack{Transit\\On-time} &
\shortstack{Transit\\Late} \\
\midrule
Drive-Early        &
\cellcolor{green!20}-0.234 & \cellcolor{green!20}0.141 & 0.011 &
0.005 & 0.013 & \cellcolor{green!20}0.002 &
0.003 & 0.009 & 0.001 &
0.003 & 0.009 & 0.000 \\
Drive-On-time       &
\cellcolor{green!20}0.034 & \cellcolor{green!20}-0.128 & \cellcolor{green!20}0.014 &
0.004 & 0.013 & \cellcolor{green!20}0.002 &
0.002 & 0.009 & 0.001 &
\cellcolor{green!20}0.003 & 0.010 & 0.000 \\
Drive-Late         &
\cellcolor{green!20}0.037 & \cellcolor{green!20}0.197 & \cellcolor{green!20}-0.316 &
0.004 & 0.014 & \cellcolor{green!20}0.003 &
0.002 & 0.009 & 0.001 &
\cellcolor{green!20}0.003 & 0.009 & 0.000 \\
Passenger-Early    &
\cellcolor{green!20}0.034 & 0.097 & 0.008 &
-0.115 & 0.017 & \cellcolor{green!20}0.003 &
0.003 & 0.011 & \cellcolor{green!20}0.002 &
0.003 & 0.011 & 0.000 \\
Passenger-On-time   &
0.024 & \cellcolor{green!20}0.102 & 0.008 &
0.005 & \cellcolor{green!20}-0.107 & \cellcolor{green!20}0.003 &
0.002 & 0.011 & \cellcolor{green!20}0.002 &
0.003 & \cellcolor{green!20}0.016 & 0.000 \\
Passenger-Late     &
0.025 & 0.102 & 0.009 &
0.005 & 0.018 & -0.116 &
0.002 & 0.011 & \cellcolor{green!20}0.002 &
0.003 & 0.011 & 0.000 \\
Ridesharing-Early  &
\cellcolor{green!20}0.032 & 0.083 & \cellcolor{green!20}0.006 &
0.006 & 0.014 & \cellcolor{green!20}0.003 &
\cellcolor{green!20}-0.248 & 0.013 & \cellcolor{green!20}0.002 &
0.004 & 0.012 & 0.000 \\
Ridesharing-On-time &
0.021 & 0.084 & 0.006 &
0.004 & 0.014 & \cellcolor{green!20}0.003 &
0.003 & -0.232 & \cellcolor{green!20}0.002 &
\cellcolor{green!20}0.004 & 0.012 & 0.000 \\
Ridesharing-Late   &
0.022 & 0.088 & 0.007 &
0.004 & 0.015 & \cellcolor{green!20}0.003 &
0.003 & 0.012 & -0.232 &
\cellcolor{green!20}0.004 & 0.012 & 0.000 \\
Transit-Early      &
0.016 & 0.056 & 0.004 &
0.003 & 0.009 & \cellcolor{green!20}0.002 &
0.002 & 0.008 & 0.001 &
-0.072 & 0.023 & 0.002 \\
Transit-On-time    &
0.013 & \cellcolor{green!20}0.060 & 0.004 &
0.003 & \cellcolor{green!20}0.014 & \cellcolor{green!20}0.002 &
0.002 & 0.008 & 0.001 &
\cellcolor{green!20}0.007 & \cellcolor{green!20}-0.058 & 0.001 \\
Transit-Late       &
0.014 & 0.060 & 0.005 &
0.003 & 0.010 & \cellcolor{green!20}0.002 &
0.002 & 0.008 & 0.001 &
\cellcolor{green!20}0.019 & 0.023 & \cellcolor{green!20}-0.089 \\
\botrule
\end{tabular*}

\end{sidewaystable}

Transit elasticities, although generally lower than ridesharing, show clear responsiveness to increased transit costs, particularly late departures. Increasing transit costs moderately pushes users to driving or ridesharing, reinforcing the need to maintain affordable transit to support sustainable mode choices.

The CNL model consistently yields stronger, more intuitive results. For instance, \textit{Drive}$_\text{Late}$ shows direct elasticity of -0.316 in CNL vs. -0.234 in MNL, suggesting that the CNL model better captures how cost-sensitive late-period drivers are.

\subsubsection{Analysis of Toll Elasticity}

Since tolls are directly applied only to Driving, Passenger (half rate), and Ridesharing (quarter rate), the Transit mode remains toll-free. Therefore, the cross-elasticities between the driving (and associated modes) and transit alternatives indicate how sensitive transit mode choices are to toll increases in driving modes.

As illustrated in Tables \ref{tab:toll_elasticity_mnl_cnl}, Direct elasticities are consistently stronger for driving alternatives, indicating that toll increases substantially reduce the attractiveness of driving, particularly during peak (On-time) and late departure times. Elasticity magnitudes for driving alternatives in the CNL model range approximately from -0.08 (Drive On-time) to -0.11 (Drive Early and Late), slightly lower than those in the MNL model. This indicates that travellers exhibit moderate responsiveness to tolling, especially for peak and late periods. Direct elasticities for ridesharing are around -0.02 to -0.03, significantly lower than driving or passenger. This lower sensitivity highlights ridesharing as a more stable alternative under congestion pricing scenarios, making it a suitable candidate to promote congestion management. Highlighted green cells in Table \ref{tab:toll_elasticity_mnl_cnl}.b denote toll elasticities that are larger in magnitude under the CNL model compared to the MNL, illustrating where the more flexible substitution structure of the CNL model results in different behavioural responses.

Positive and modest cross elasticity values across alternatives demonstrate potential substitution effects, while increasing tolls for driving (early, on-time, late) consistently results in positive cross-elasticities for the transit modes. This indicates travellers consider transit a viable alternative when driving costs increase due to tolls.

\begin{sidewaystable}
\caption{Toll elasticity estimates for the MNL and CNL models}
\label{tab:toll_elasticity_mnl_cnl}
\footnotesize
\renewcommand{\arraystretch}{1}

% =============== (a) MNL ==================
\begin{tabular*}{\textheight}{@{\extracolsep\fill}lrrrrrrrrr}
\toprule
\multicolumn{10}{l}{\textbf{(a) Toll Elasticity for the MNL Model}}\\[4pt]
\textbf{Alternative / Variable} &
\shortstack{Drive\\Toll\\Early} &
\shortstack{Drive\\Toll\\On-time} &
\shortstack{Drive\\Toll\\Late} &
\shortstack{Pass\\Toll\\Early} &
\shortstack{Pass\\Toll\\On-time} &
\shortstack{Pass\\Toll\\Late} &
\shortstack{Rideshare\\Toll\\Early} &
\shortstack{Rideshare\\Toll\\On-time} &
\shortstack{Rideshare\\Toll\\Late} \\
\midrule
Drive-Early        & -0.124 & 0.086 & 0.004 & 0.003 & 0.012 & 0.001 & 0.000 & 0.001 & 0.000 \\
Drive-On-time      &  0.017 & -0.101 & 0.004 & 0.002 & 0.012 & 0.001 & 0.000 & 0.001 & 0.000 \\
Drive-Late         &  0.020 & 0.098 & -0.101 & 0.003 & 0.014 & 0.001 & 0.000 & 0.001 & 0.000 \\
Passenger-Early    &  0.019 & 0.082 & 0.004 & -0.075 & 0.016 & 0.001 & 0.000 & 0.002 & 0.000 \\
Passenger-On-time  &  0.017 & 0.083 & 0.004 & 0.003 & -0.087 & 0.001 & 0.000 & 0.002 & 0.000 \\
Passenger-Late     &  0.018 & 0.091 & 0.005 & 0.004 & 0.018 & -0.056 & 0.000 & 0.002 & 0.000 \\
Ridesharing-Early  &  0.017 & 0.072 & 0.003 & 0.003 & 0.014 & 0.001 & -0.027 & 0.002 & 0.000 \\
Ridesharing-On-time&  0.015 & 0.074 & 0.004 & 0.003 & 0.014 & 0.001 & 0.000 & -0.035 & 0.000 \\
Ridesharing-Late   &  0.016 & 0.079 & 0.004 & 0.003 & 0.016 & 0.001 & 0.000 & 0.002 & -0.020 \\
Transit-Early      &  0.011 & 0.051 & 0.003 & 0.002 & 0.010 & 0.000 & 0.000 & 0.001 & 0.000 \\
Transit-On-time    &  0.010 & 0.052 & 0.003 & 0.002 & 0.010 & 0.000 & 0.000 & 0.001 & 0.000 \\
Transit-Late       &  0.011 & 0.055 & 0.003 & 0.002 & 0.011 & 0.001 & 0.000 & 0.001 & 0.000 \\
\botrule
\end{tabular*}

\vspace{1cm}

% =============== (b) CNL ==================
\begin{tabular*}{\textheight}{@{\extracolsep\fill}lrrrrrrrrr}
\toprule
\multicolumn{10}{l}{\textbf{(b) Toll Elasticity for the CNL Model}}\\[4pt]
\textbf{Alternative / Variable} &
\shortstack{Drive\\Toll\\Early} &
\shortstack{Drive\\Toll\\On-time} &
\shortstack{Drive\\Toll\\Late} &
\shortstack{Pass\\Toll\\Early} &
\shortstack{Pass\\Toll\\On-time} &
\shortstack{Pass\\Toll\\Late} &
\shortstack{Rideshare\\Toll\\Early} &
\shortstack{Rideshare\\Toll\\On-time} &
\shortstack{Rideshare\\Toll\\Late} \\
\midrule
Drive-Early         & -0.112 & \cellcolor{green!20}0.087 & 0.004 & 0.003 & 0.008 & 0.001 & 0.000 & 0.001 & 0.000 \\
Drive-On-time       &  0.016 & -0.084 & \cellcolor{green!20}0.005 & 0.002 & 0.009 & 0.001 & 0.000 & 0.001 & 0.000 \\
Drive-Late          &  0.019 & \cellcolor{green!20}0.132 & \cellcolor{green!20}-0.108 & 0.002 & 0.010 & 0.001 & 0.000 & 0.001 & 0.000 \\
Passenger-Early     &  0.017 & 0.062 & 0.003 & -0.058 & 0.011 & 0.001 & 0.000 & 0.002 & 0.000 \\
Passenger-On-time   &  0.012 & 0.066 & 0.003 & 0.002 & -0.071 & 0.001 & 0.000 & 0.002 & 0.000 \\
Passenger-Late      &  0.012 & 0.068 & 0.003 & 0.003 & 0.012 & -0.040 & 0.000 & 0.002 & 0.000 \\
Ridesharing-Early   &  0.017 & 0.055 & 0.002 & 0.003 & 0.010 & 0.001 & \cellcolor{green!20}-0.028 & 0.002 & 0.000 \\
Ridesharing-On-time &  0.011 & 0.058 & 0.002 & 0.002 & 0.010 & 0.001 & 0.000 & -0.034 & 0.000 \\
Ridesharing-Late    &  0.011 & 0.060 & 0.003 & 0.002 & 0.011 & 0.001 & 0.000 & 0.002 & -0.019 \\
Transit-Early       &  0.008 & 0.039 & 0.002 & 0.002 & 0.007 & 0.001 & 0.000 & 0.001 & 0.000 \\
Transit-On-time     &  0.007 & 0.042 & 0.002 & 0.001 & 0.009 & \cellcolor{green!20}0.001 & 0.000 & 0.001 & 0.000 \\
Transit-Late        &  0.007 & 0.043 & 0.002 & 0.002 & 0.007 & 0.001 & 0.000 & 0.001 & 0.000 \\
\botrule
\end{tabular*}

\end{sidewaystable}

Toll elasticities, critical for congestion pricing, reinforce the CNL’s advantage. Direct elasticities for toll-sensitive alternatives like \textit{Drive}$_\text{Late}$ are stronger in the CNL (-0.108) than in the MNL (-0.101). Similarly, cross-elasticities for toll adjustments in passenger and ridesharing modes indicate greater responsiveness in the CNL model, providing more nuanced insights for policy making.

\subsubsection{Policy Implications for Congestion Pricing}

The elasticity analysis provides valuable insights for designing effective congestion pricing policies. High direct time elasticity for late-driving commuters indicates a strong potential to reduce peak-hour congestion by implementing higher tolls during these periods. This could incentivize a shift to earlier or on-time departures or even a mode change. Similarly, moderate elasticity observed for early departures suggests that modest incentives such as off-peak discounts, could further encourage travel during less congested times, helping to distribute demand more evenly across the day.

Drivers demonstrate considerable within-mode flexibility, as seen in the high time cross-elasticities across departure periods. Dynamic tolling strategies that vary by time of day could capitalize on this flexibility to balance traffic loads, smoothing peak-hour spikes. While cross-modal substitution is more limited, the elasticity patterns underscore the importance of enhancing transit travel time reliability and comfort to strengthen public transit's competitiveness relative to private vehicles.

Ridesharing and passenger modes show high sensitivity to changes in travel time, highlighting the importance of maintaining efficient and time-competitive mobility options. If congestion pricing policies lead to longer travel times or perceived inconvenience in shared modes, their attractiveness may diminish. Therefore, it is essential to integrate complementary measures, such as priority lanes or real-time updates, especially for shared and high-occupancy vehicles. Transit users were also highly time-sensitive, reinforcing the recommendation to channel congestion pricing revenues into improvements in service frequency and capacity to both retain current users and attract new ones.

The differentiation in elasticity magnitudes by both mode and departure time supports the case for a hybrid congestion pricing scheme. Combining spatial (cordon-based), temporal (time-of-day), and modal pricing elements allows for precision targeting, charging where and when elasticity is higher to achieve greater behavioral change. For example, cost elasticity is particularly pronounced for driving during late and on-time departures, reinforcing the effectiveness of peak-period pricing strategies to encourage earlier departures or reduce overall demand.

Ridesharing users exhibit strong cost sensitivity, indicating that price increases may prompt shifts back to single-occupancy driving if not carefully managed. Policymakers should exercise caution in adjusting ridesharing fares and consider complementary policies such as shared ride incentives or infrastructure improvements. Transit fares, meanwhile, should remain stable or subsidized during peak periods to prevent cost-induced shifts back to private vehicles.

Given the central concentration of destinations with high cost and time elasticities, cordon-based tolling targeting the downtown core offers a strategic opportunity to influence travel choices where impacts are greatest. Such spatial pricing, when paired with investments in public transit and early-departure incentives, can shift behavior meaningfully without excessive burden on users.
The findings also raise equity concerns, as passenger and transit users, who may include lower-income or carless individuals, exhibit higher cost elasticities. To mitigate potential adverse effects, pricing policies should include exemptions or discounts for essential travel such as medical, educational, or caregiving trips.

Finally, toll elasticities provide further policy direction. Lower elasticity among on-time drivers implies they are less likely to shift mode or departure time, making them a reliable revenue source during peak hours. Targeted tolls during these periods can effectively manage congestion with minimal disruption. Conversely, somewhat higher toll elasticities for early and late departures suggest these groups are more responsive to pricing. Offering reduced tolls during these periods could help redistribute demand away from the peak. Ridesharing users, facing only partial toll rates, show relatively low direct toll sensitivity. This presents a valuable opportunity to promote shared modes through continued toll discounts, priority infrastructure, or service enhancements.

In sum, Congestion pricing, particularly if targeted by travel time and location, has strong potential to reduce congestion and promote mode shifts, especially toward public transit. Complementing these strategies with reinvestment in transit infrastructure and operational quality will be essential to ensure behavioral changes are both substantial and sustainable.

\subsection{Spatial Elasticity Analysis}

\begin{figure}[h]
    \centering
    \includegraphics[width=0.9\textwidth]{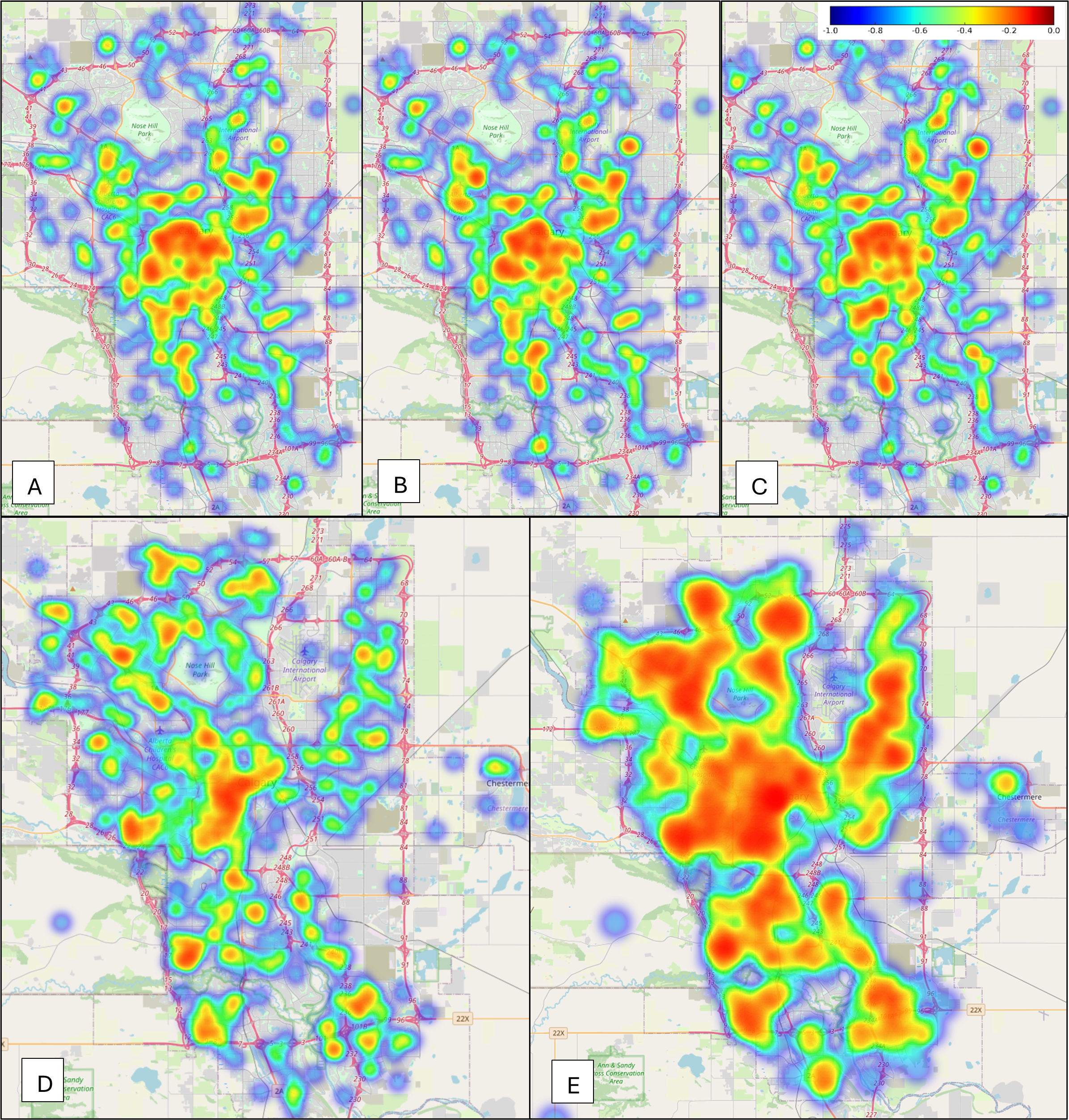}
    \caption{Time Direct Elasticity Heatmaps: \textbf{A)} Drive Early vs Destination, \textbf{B)} Drive Late vs Destination, \textbf{C)} Ridesharing On time vs Destination, \textbf{D)} Drive On time vs Origin, \textbf{E)} Transit On time vs Origin. \\
    Red areas indicate higher (stronger negative) elasticity values, meaning greater sensitivity to increased travel time, while blue areas represent lower sensitivity.}
    \label{fig:time_elasticity_heatmaps}
\end{figure}

The spatial heatmaps of direct travel time elasticities derived from the CNL model (Figure \ref{fig:time_elasticity_heatmaps}) highlight distinct geographic patterns in travellers’ responsiveness to changes in travel time across Calgary. Drive–Early elasticities are strongly negative and concentrated in the downtown and Beltline areas, indicating that early car commuters accessing central employment zones are particularly time-sensitive. In contrast, Drive–On-time elasticities are more pronounced in suburban residential neighbourhoods, especially across the northeast and southwest quadrants, suggesting greater responsiveness to travel delays among typical peak-hour commuters in these peripheral zones. Drive–Late elasticities exhibit moderate magnitudes and are concentrated along key urban arteries and corridors serving hospitals and large institutions, reflecting potential for modest behavioural shifts among late-period drivers, including those engaged in shift-based or flexible employment. These findings support the strategic use of targeted temporal pricing schemes, such as peak tolls or early incentives, tailored to the travel patterns of both suburban and centrally bound commuters.

Ridesharing, as seen in the Ridesharing–On-time destination elasticity map, exhibits lower but still spatially relevant sensitivity, primarily clustered around the city centre. This points to opportunities for targeted pricing incentives that favour shared mobility during congested periods. Collectively, the elasticity heatmaps underscore the viability of spatially and temporally differentiated congestion pricing strategies. For transit users departing on time, the elasticity maps for origins highlight notable sensitivity to travel time increases in areas like the northeast and southern parts of the city, where access to frequent, reliable transit remains essential. While the southeast contains more industrial and employment zones, the transit sensitivity there may reflect workers’ dependence on scheduled services. On the other hand, higher residential areas in the southwest show moderate magnitude but consistent sensitivity, suggesting an opportunity to enhance service speed and reliability in these corridors.

\begin{figure}[h]
    \centering
    \includegraphics[width=0.9\textwidth]{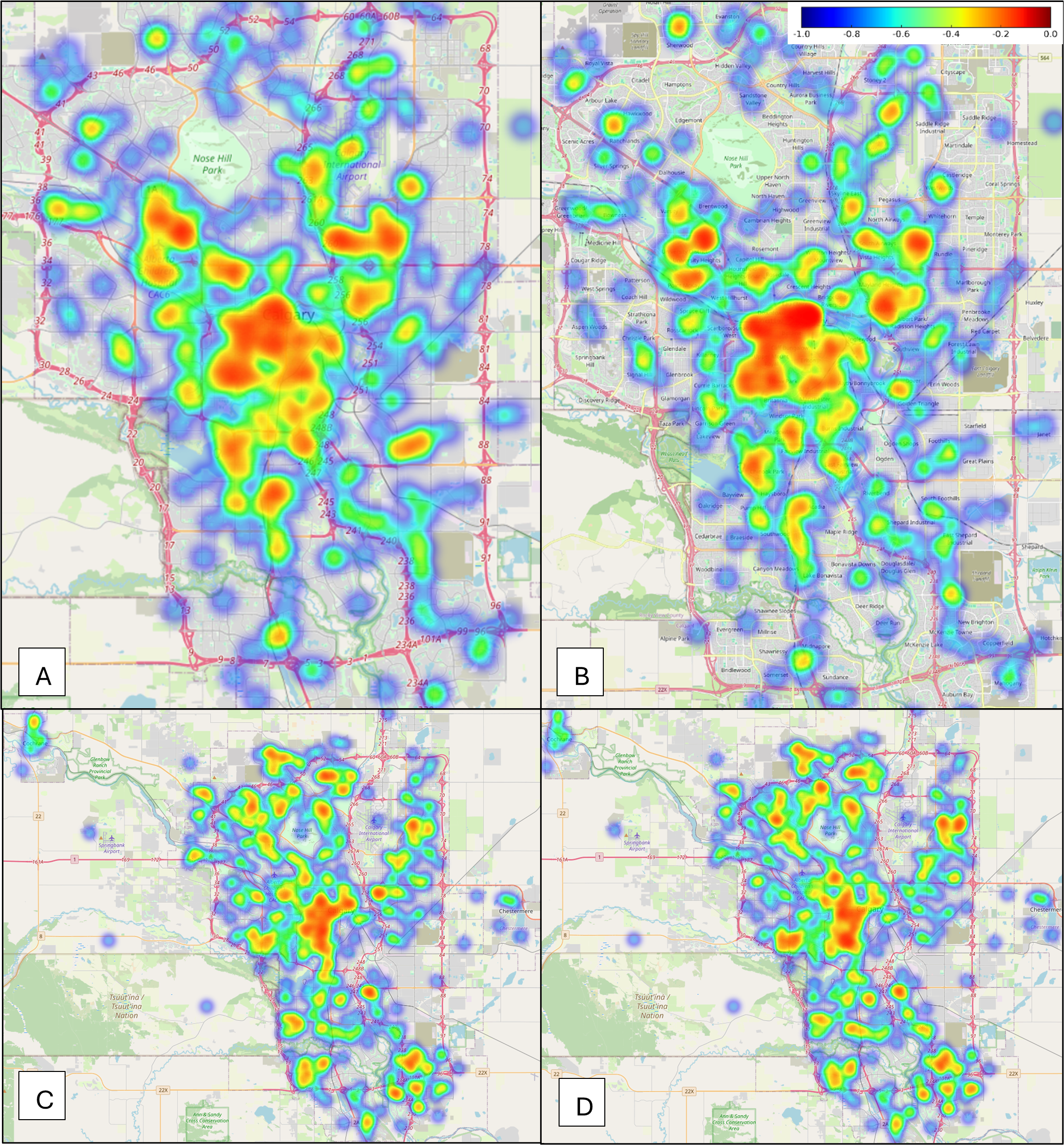}
    \caption{Cost Direct Elasticity Heatmaps: \textbf{A)} Drive On time vs Destination, \textbf{B)} Transit On time vs Destination, 
    \textbf{C)} Drive Early vs Origin, \textbf{D)} Ridesharing On time vs Origin. \\
    Red areas indicate higher (stronger negative) elasticity values, meaning greater sensitivity to increased travel cost, while blue areas represent lower sensitivity.}
    \label{fig:cost_elasticity_heatmaps}
\end{figure}

The cost elasticity heatmaps (Figure \ref{fig:cost_elasticity_heatmaps}) reveal distinct spatial and temporal sensitivities to increased travel costs, particularly for driving during peak periods. Among the most insightful are the maps for Drive On-time vs destination, which show notable reductions in travel demand concentrated in the downtown core and surrounding employment hubs. These results suggest that even a modest cost increase (e.g., through tolling or parking fees or gas price) during peak hours could meaningfully discourage car use in the most congested zones of the city. The southwest and northeast quadrants, key origin points for daily commuting, also exhibit elevated cost sensitivity, indicating that many long-distance commuters from these areas are responsive to pricing signals.

In contrast, the elasticity maps for Drive Early and Transit On-time show much lower sensitivity across the city, suggesting that early car trips and scheduled transit-based travel are less responsive to cost-based interventions. The Ridesharing On-time heatmap (by origin) reveals that increasing the travel costs for ridesharing would likely reduce the attractiveness of this option for many residents, particularly in central and northern Calgary. While the model does not include trip cancellation as an option, the results indicate a potential shift away from ridesharing toward alternative modes or departure times, rather than a reduction in total travel demand. The intensity of the response in denser neighborhoods implies that such pricing could effectively shift travel behavior, potentially encouraging mode shifts or temporal adjustments, particularly if viable alternatives such as transit are readily accessible. These findings support a targeted policy approach: tolling strategies should prioritize peak period driving into the downtown core while ensuring that transit remains a viable, affordable alternative. The observed variation by departure time also highlights the importance of temporal granularity in pricing schemes, reinforcing the value of time-based tolling or rebate structures to shift demand and alleviate pressure on the transportation network during the peak-hour congestion.

\begin{figure}[h]
    \centering
    \includegraphics[width=\textwidth]{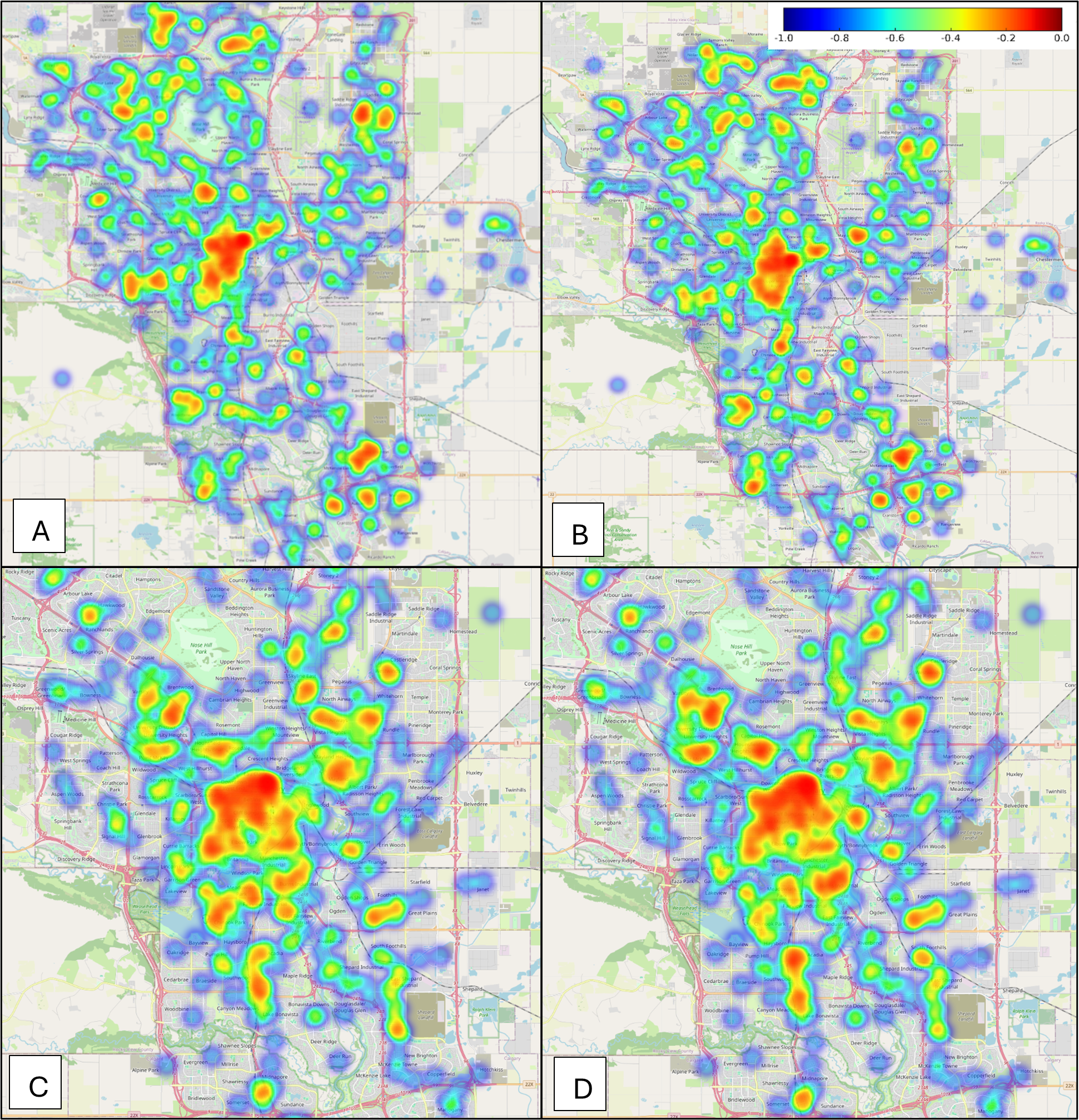}
    \caption{Toll Direct Elasticity Heatmaps: \textbf{A)} Drive On time vs Origin, \textbf{B)} Drive Early vs Origin, 
    \textbf{C)} Drive On time vs Destination, \textbf{D)} Drive Late vs Destination. \\
    Red areas indicate higher (stronger negative) elasticity values, meaning greater sensitivity to increased toll, while blue areas represent lower sensitivity.}
    \label{fig:toll_elasticity_heatmaps}
\end{figure}

Reviewing the spatial distribution of direct toll elasticities heatmaps (Figure \ref{fig:toll_elasticity_heatmaps}), Notably, the destination-based elasticity for Drive On-time trips, which shows the highest negative response concentrated in Calgary's downtown and Beltline areas, indicates that commuters arriving at these central employment hubs during peak hours are particularly responsive to toll increases, likely due to a combination of higher congestion, greater availability of modal alternatives, and workplace flexibility that allows schedule adjustments. These findings support the application of a cordon-based pricing scheme focused on peak-period access to downtown, where demand reduction is both needed and feasible.

In contrast, the origin-based toll elasticity for Drive On-time exhibits lower responsiveness, especially in peripheral neighborhoods such as Airdrie, Cochrane, and southeast Calgary. These areas tend to be more car-dependent due to limited access to regional transit services, longer commute distances, and fewer modal alternatives, making toll-induced behavioural shifts less likely without additional intervention. Such interventions could include expanding park-and-ride facilities, enhancing feeder bus networks, or providing targeted transit subsidies to suburban communities. The Drive Early origin heatmap further reinforces this insight by showing diffuse, modest sensitivities across residential suburbs, suggesting that early commuters may lack schedule flexibility or viable alternatives. Meanwhile, the destination elasticity for Drive Late trips illustrates another opportunity: significant responsiveness among those arriving downtown just outside peak hours. This implies that marginal toll discounts or incentives for early or late arrivals could help shift demand out of the most congested periods, flattening the peak and improving network performance. Together, these heatmaps demonstrate the need for a targeted and inclusive tolling approach, leveraging spatial and temporal variation in elasticity to optimize system-wide outcomes while safeguarding access for lower-mobility populations.

To further explore substitution effects under congestion pricing, four destination-based cross-elasticity heatmaps (Drive On-time vs Drive Early, Drive Late, Ridesharing On-time and Transit On-time) were analyzed to assess how changes in tolls influence the attractiveness of alternative travel choices. The spatial distribution of these elasticities reveals that destinations within Calgary's central core, particularly employment-dense downtown zones, exhibit higher sensitivity to toll changes, likely due to the greater availability of viable alternatives such as robust transit services and walkable access. For instance, individuals commuting to central destinations are likely to substitute away from tolled Drive On-Time options in favour of alternatives, such as earlier departures or public transit. Similarly, moderate elasticities in educational and mixed-use areas, reflected in Drive Late Heatmap, highlight the potential for tolling to influence travel decisions where moderate congestion and schedule flexibility coexist.

\begin{figure}[h]
    \centering
    \includegraphics[width=\textwidth]{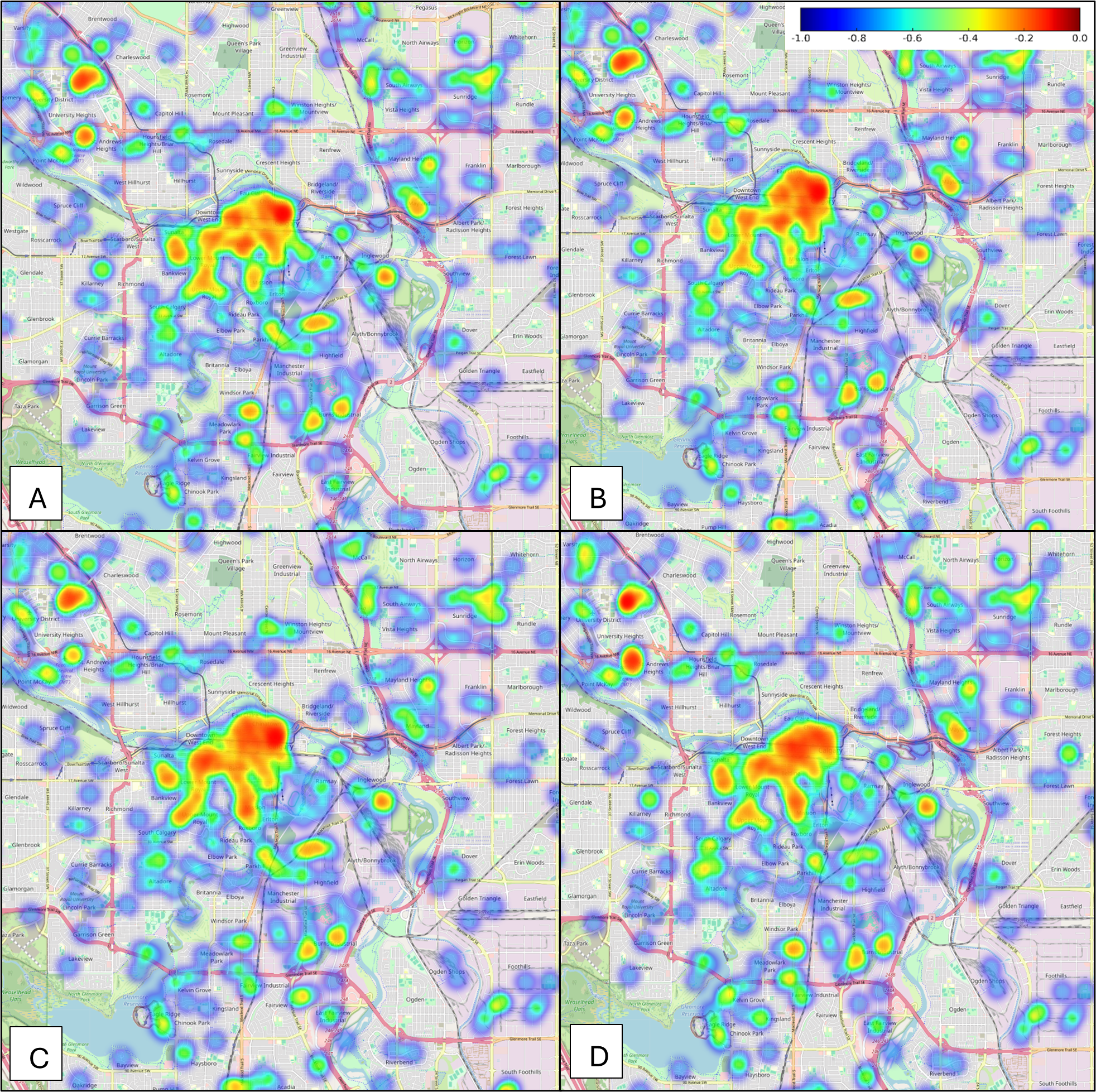}
    \caption{Toll Cross Elasticity Heatmaps for change in toll of Drive On time affecting destination vs. 
    \textbf{A)} Drive Early, \textbf{B)} Drive Late, \textbf{C)} Ridesharing On time, \textbf{D)} Transit On time. \\
    Red areas indicate higher (stronger negative) elasticity values, meaning greater sensitivity to increased toll, while blue areas represent lower sensitivity.}
    \label{fig:toll_cross_elasticity_heatmaps}
\end{figure}

Ridesharing and Transit heatmaps further reveal important dynamics in suburban and peripheral employment zones. These areas display varying levels of cross-elasticity, indicating that commuters headed to light industrial or retail destinations may shift travel times or modes in response to toll increases, particularly where flexible scheduling or shared mobility options are available. The results underscore the spatial heterogeneity of behavioral responses: while downtown pricing may achieve strong deterrence effects, outer zones may require complementary measures such as improved transit, employer-based incentives, or dynamic tolling to induce meaningful changes. Overall, integrating cross-elasticity insights into pricing strategies enhances the capacity to design equitable and effective congestion pricing policies that account for spatial diversity in traveler responsiveness.

Further analysis has been done to study how toll elasticity varies across both departure times and travel modes. When comparing toll elasticities for driving across early, on-time, and late departures, on-time drivers consistently show the lowest sensitivity to tolls. This pattern is especially clear across all trip distances, reinforcing that commuters traveling during peak periods are less responsive to price signals, likely due to rigid schedules. In contrast, early and late drivers demonstrate higher elasticities, particularly for short and medium-length trips. This suggests they are more likely to adjust their travel time in response to toll changes. The greater spread of elasticity values among early and late drivers, especially at longer distances, points to more variation in flexibility or access to alternatives.

Comparing toll elasticities across modes for on-time departures, shows a marked contrast. Ridesharing trips exhibit minimal responsiveness to toll changes, likely because costs are shared or absorbed. Passenger modes show slightly more sensitivity, while on-time solo drivers remain the most responsive among the three. These patterns highlight the stability of ridesharing under tolling and suggest it could be strategically promoted as a viable alternative to driving alone under a road pricing regime. Altogether, the graphs emphasize that both departure time and travel mode matter when designing tolling schemes, and that travelers’ sensitivity varies not just by price, but also by their trip purpose, flexibility, and distance.

On-time drivers, who exhibit the lowest toll elasticities, are least likely to shift behavior, making them suitable targets for higher tolls during peak periods, especially for trips into central areas. Because they are less price-sensitive, these users can generate reliable revenue without major behavioral shifts, supporting system improvements elsewhere. In contrast, early and late drivers are more responsive to tolls, particularly over shorter distances. Moderate tolls during these shoulder periods could effectively spread peak demand, encouraging commuters with more flexible schedules to adjust their departure time and relieve pressure on the system during peak congestion windows.

The comparison across modes for on-time departures reveals another layer of insight. Ridesharing users are the least sensitive to toll changes, suggesting that small toll adjustments won’t deter their use. This provides an opportunity to promote ridesharing as a stable alternative to solo driving, particularly if supported by dedicated curb access or HOV lanes. Passenger modes are somewhat more elastic, implying that even modest toll increases could discourage use unless paired with offsetting incentives, again suggesting HOV lanes are an important policy to consider in partnership with road pricing. 

\section{Conclusion and Future Works}\label{sec5}

The analyses conducted in this paper demonstrate that the CNL model outperforms traditional MNL and NL models by capturing the nuanced and multidimensional responses of travellers to congestion pricing policies. By estimating a joint mode and departure time choice model that allows for flexible substitution patterns across both dimensions, this study provides a more behaviourally realistic understanding of how commuters respond to various pricing strategies. Through detailed discrete choice modelling, spatial analysis, and elasticity assessments, congestion pricing emerges as a highly effective tool for influencing traveller behaviour, encouraging shifts in both mode and departure time. Such behavioural adjustments can significantly reduce urban traffic congestion.

Travel time, cost, and toll elasticities reveal significant traveller sensitivity, particularly among peak-hour commuters and central-area travellers. High elasticity during late departures and substantial cross-elasticities between driving departure times suggest that dynamic tolling strategies can effectively redistribute peak-period demand. Spatial elasticity heatmaps further underline this finding, showing high responsiveness to toll increases for central destinations and comparatively lower responsiveness in peripheral residential areas, highlighting the importance of geographically differentiated pricing strategies.

This paper contributes to the literature in several key ways. First, it develops and estimates a comprehensive CNL model that jointly accounts for mode and departure time choice under multiple pricing schemes, including cordon-based, distance-based, and travel-time-based tolls. Second, it incorporates attitudinal variables and weather conditions, offering richer insights into contextual and environmental influences on commuter behaviour. Third, by integrating spatial and cross-elasticity analyses, the study advances a geographically and temporally disaggregated perspective on pricing impacts, critical for real-world policy implementation. Finally, the inclusion of socio-demographic heterogeneity provides empirical evidence on income-related sensitivities to tolling, supporting more informed policy design.

Travel time, cost, and toll elasticities reveal significant traveller sensitivity, particularly among peak-hour commuters and central-area travellers. High elasticity during late departures and substantial cross-elasticities between driving departure times suggest that dynamic tolling strategies can effectively redistribute peak-period demand. Spatial elasticity heatmaps further underline this finding, showing high responsiveness to toll increases for central destinations and comparatively lower responsiveness in peripheral residential areas, highlighting the importance of geographically differentiated pricing strategies. further analyses emphasize that commuters show the least flexibility during on-time peak departures, suggesting that this group can be targeted with higher tolls to regulate congestion without significantly disrupting travel behaviour. In contrast, early and late drivers, as well as ridesharing users, demonstrate higher elasticity, presenting opportunities to encourage off-peak and shared travel through moderate pricing and supportive incentives. These insights support the adoption of a hybrid congestion pricing strategy that blends multiple schemes, cordon-based charges to regulate central congestion, distance-based pricing to reflect network usage, and travel-time-based tolling to address temporal demand pressures. This appears most effective when complemented by transit and ridesharing improvements. To ensure both public acceptability and policy sustainability, future strategies should consider incorporating targeted exemptions or reduced rates for vulnerable travellers. This integrated approach mirrors recent initiatives such as New York City’s congestion pricing plan, which combines spatial and temporal pricing elements to manage urban traffic \citep{cook2025short}. A similarly layered strategy in Calgary could enhance responsiveness, distribute burdens more fairly, and improve policy salience and public acceptability.

Like all stated preference studies, this research has certain limitations. The behavioural responses observed are based on hypothetical scenarios rather than revealed choices, which may introduce bias due to hypothetical or strategic response effects. Additionally, while the CNL model offers improved behavioural realism, it still relies on parametric assumptions that constrain substitution patterns to some extent. Future research could advance this work by applying hybrid choice models that integrate attitudinal or psychological factors, such as perceptions of fairness, environmental concern, or altruism, into the behavioural framework, thereby capturing a deeper layer of decision-making complexity. Moreover, incorporating latent class models or mixed logit formulations can help account for unobserved heterogeneity across user segments. The use of real-time data from smart mobility systems or revealed preference datasets would also enable validation and refinement of the findings. Finally, given the distributive implications of pricing policies, future studies should explicitly examine equity impacts by developing fairness-based modelling frameworks and evaluating policy outcomes across sociodemographic and spatial dimensions.



\end{document}